\documentclass[a4paper,10pt]{elsarticle}
\usepackage{amsmath,bm,graphicx}
\usepackage{subfigure}
\usepackage{amssymb}
\usepackage{amsfonts}

\newcommand{\eref}[1]{{(\ref{#1})}}
\newcommand{\mbf}[1]{{\mathbf{#1}}}
\begin{document}
\begin{frontmatter}

\title{Complete classification of discrete resonant Rossby/drift wave triads on periodic domains}

\author{Miguel D. Bustamante}
%\email{miguel.bustamante@ucd.ie}
\author{Umar Hayat}
%\email{umarmaths@gmail.com}

\address{{School of Mathematical Sciences, University College Dublin, Belfield, Dublin 4, Ireland}}
\begin{abstract}
We consider the set of Diophantine equations that arise in the context of the partial differential equation called ``barotropic vorticity equation'' on periodic domains, when nonlinear wave interactions are studied to leading order in the amplitudes. The solutions to this set of Diophantine equations are of interest in atmosphere (Rossby waves) and Tokamak plasmas (drift waves), because they provide the values of the spectral wavevectors that interact resonantly via three-wave interactions. These wavenumbers come in ``triads'', i.e., groups of three wavevectors.

We provide the full solution to the Diophantine equations in the physically sensible limit when the Rossby deformation radius is infinite. The method is completely new, and relies on mapping the unknown variables via rational transformations, first to rational points on elliptic curves and surfaces, and from there to rational points on quadratic forms of ``Minkowski'' type (such as the familiar \emph{space-time} in special relativity). Classical methods invented centuries ago by Fermat, Euler, Lagrange, Minkowski, are used to classify all solutions to our original Diophantine equations, thus providing a computational method to generate numerically all the resonant triads in the system. Computationally speaking, our method has a clear advantage over brute-force numerical search: on a $10000^2$ grid, the brute-force search would take 15 years using optimised C${}^{++}$ codes on a cluster, whereas our method takes about 40 minutes using a laptop.

Moreover, the method is extended to generate so-called quasi-resonant triads, which are defined by relaxing the resonant condition on the frequencies, allowing for a small mismatch. Quasi-resonant triads' distribution in wavevector space is robust with respect to physical perturbations, unlike resonant triads' distribution. Therefore, the extended method is really valuable in practical terms. We show that the set of quasi-resonant triads form an intricate network of connected triads, forming clusters whose structure depends on the value of the allowed mismatch. It is believed that understanding this network is absolutely relevant to understanding turbulence. We provide some quantitative comparison between the clusters' structure and the onset of fully nonlinear turbulent regime in the barotropic vorticity equation, and we provide perspectives for new research.

\end{abstract}

\end{frontmatter}

\section{Introduction}
\label{sec-intro}

Understanding the behaviour of nonlinear wave resonances is of importance in numerical weather prediction, fusion reactors and in general in experiments in fibre optics and water waves. Modulational instability is the first example of relevance of resonant and quasi-resonant triads and quartets \cite{Be67}.  Effects such as wave turbulence with its cascades of energy and enstrophy rely on the interconnected web of resonances.  However a thorough understanding of the full-scale mechanism of energy transfers is lacking at the moment and one of the reasons for this is the poor understanding of the web of resonant and quasi-resonant triads. This poor understanding is in two aspects: (i) Kinematical aspect, regarding the actual position in wavenumber space of the modes that interact in resonant and quasi-resonant triads, as well as the interconnections of these triads to form clusters. (ii) Dynamical aspect, regarding the processes of energy exchanges between modes via $3$-wave interactions. This paper will concentrate on the kinematical aspect, in the context of the barotropic vorticity equation on periodic domains. As a motivation, we believe that the information obtained from the kinematical aspects can be used to derive a reduced approximated system of evolution equations for the Fourier amplitudes of the underlying PDEs that describe the nonlinear oscillations. The long-term goal is to produce an important degree of simplification from an infinite-dimensional space of dynamical variables to a finite-dimensional one, while keeping enough complexity so as to accommodate physically-measurable characteristics such as turbulence and energy/enstrophy cascades. This paradigm is in the spirit of the study of mesoscopic and discrete wave turbulence (areas under development). See for example \cite{Re89, Co01, Za05, Ka06}.

Another question is whether energy transfers between wave modes depend or not on the nature and strength of the forcing. Such question has been studied in a myriad of papers using numerical simulations of full PDE models, and the intuition gained is quite empirical. However, the understanding of the network of connected triads might lead to a better modelling of this and other problems. In \cite{Ha12}, the effect of forcing a single triad was studied and it was established that the triad's energy remains bounded, even if the forcing oscillates in phase with the natural frequency of the so-called unstable mode.  Going beyond triads, to quartets and quintets, formation of Kolmogorov-like spectra of inverse cascade of weak turbulence was observed in \cite{An96}, in a direct numerical simulation of a special formulation of water waves. This formulation, known as integro-differential Zakharov equation, was exploited numerically in the cited reference so that only a relatively small set of wave modes needed to be integrated forward in time.

The importance of the quasi-resonant triads over the exactly resonant ones for the accurate and faithful description of nonlinear physical wave systems has been verified in several works, starting from the modulational instability \cite{Be67} and nowadays it is an accepted fact that any proper description of wave interactions must take into account quasi-resonant triads from the start \cite{Colm,Smith2005,Alam}. See also \cite{Janssen,An2006} for analogous results in the context of near-resonant quartets and quintets.

In Jupiter's atmosphere, local wave oscillations have been observed with wavelengths of 20 km \footnote{http://photojournal.jpl.nasa.gov/catalog/PIA00724}. If such waves could be extended to cover a significant part of Jupiter's surface then we would have wavenumbers of approximately $3000$. In numerical simulations, particularly in atmospheric models, it has been shown \cite{Drit} that high resolutions are needed if one wants to describe properly the generation of small-scale turbulence out of a large-scale initial condition (see the benchmark problem of Rossby-Haurwitz wave in the cited reference). Therefore, running $2$D simulations with up to $10000$ grid points per spatial direction is meaningful, particularly if one is interested in studying inverse energy cascades. The problem with these high-resolution simulations, is that a rigorous analysis of the wave interactions (particularly in the limit of very small amplitudes, used in wave turbulence) escapes from the computational capabilities of the current computer technology. For example, if we try to do a numerical search of the quasi-resonant modes with wavenumbers less than or equal to $10000$ in size, a computer will need to do a search over $10000^4$ possible triads, and evaluate the frequency mismatch. If $\tau$ is the time it takes the computer to do one triad, then the total time for this search will be approximately $\tau \times 10^{16}.$ On a medium-sized cluster, using optimised $C^{++}$ codes, the effective time for one such computation is about $\tau \approx 5 \times 10^{-8}$ seconds, which means that the total computing time for all triads is about $15$ years!

The search for quasi-resonances becomes therefore a computational problem and this paper provides a solution, with a computational method that is based on a mapping from resonant triads to rational points on quadratic hyper-surfaces of ``Minkowski'' type (i.e., non-definite). The method for searching exact resonances is extended to a method for finding quasi-resonant triads. Here the advantage of the method over brute-force direct search is that the method produces quasi-resonant triads with preferentially small detuning levels, which is useful physically, since we are interested in the regime of small enough amplitudes. A \emph{Mathematica} code that generates the triads using this method is available from the authors upon request.

The paper is organised as follows. In Section \ref{sec:bve} we review the barotropic vorticity equation and the Rossby triads on a periodic spatial domain (torus). In Section \ref{sec:discrete} we state the Diophantine equation that determines the triad resonance conditions, develop the mapping to elliptic curves and to Minkowski-type of quadratic forms, establish the general solution to the resonance conditions, and explain in detail the computational method of solution, with a step-by-step formulation. In Section \ref{sec:num_res}, we present the numerical application of the method, showing how to construct the triads in the box of size $5000$ in wavenumber space, and studying the structure, distribution and connectivity properties of the resonant clusters that form. In Section \ref{sec:quasi} we introduce quasi-resonant triads, which are far more relevant physically than resonant triads. We explain the method to construct these quasi-resonant triads, which is based on the previous method of construction of resonant triads. In Section \ref{sec:new_dev} we discuss further extensions of the method to the case when the aspect ratio of the spatial variables is not equal to one. Finally, in Section \ref{sec:concl} we discuss the applicability of the method to the general case of the barotropic vorticity equation with finite Rossby deformation radius. The Appendix contains all necessary mathematical results to understand the method of construction of resonant triads presented in Section \ref{sec:discrete}.

\section{Governing equations}
\label{sec:bve}

\subsection{Rossby Triads}

We consider the dynamics of a rotating shallow layer of incompressible fluid, in the so-called $\beta$-plane approximation. The governing equation reduces to the following partial differential equation, known as the barotropic vorticity equation:
\begin{equation}
\label{eq:CHM}
 \frac{\partial}{\partial t} \left(\nabla^2 \psi - F \psi \right) + \left(\frac{\partial \psi}{\partial x} \frac{\partial \nabla^2 \psi}{\partial y}-\frac{\partial \psi}{\partial y} \frac{\partial \nabla^2 \psi}{\partial x} \right) + \beta \frac{\partial \psi}{\partial x} = 0,
\end{equation}
where $\psi = \psi(x,y,t) \in \mathbb{R}$ is the streamfunction, $\beta$ is a constant that determines the speed of rotation of the system, and $F = 1/R^2$ where $R$ is the Rossby deformation radius. For technical reasons (to be discussed in Section \ref{sec:concl}), we consider from here on the physically sensible limit of infinite Rossby deformation radius, $F=0.$ In terms of plasma notation, the physical limit that we consider in this paper is $\rho \, |\mbf{k}| \gg 1$, where $\rho$ is the Larmor radius and $\mbf{k}$ is the wavevector of the oscillations. This means that we are looking at small-scale structures (much smaller than the Larmor radius). We remark that the triads we will find still have a range of applicability into the more realistic case $F > 0$ (the explanation of how this is done is given in Section \ref{sec:concl}).

%Commented reference Rossby Radius of Deformation:
%http://taylor.math.ualberta.ca/~bruce/glossary.html

Equation (\ref{eq:CHM}) is also known in the literature as the Charney-Hasegawa-Mima equation, or CHM. The ``C'' comes from the atmospherical context just described. The ``HM'' comes from an independent derivation of the same mathematical equation in the context of plasma physics \cite{Ha78}. Due to this multidisciplinary aspect, the CHM equation occupies a special place in physics and mathematics. On the one hand, it suffices to say that there is no general solution of the equation. Due to its nonlinear term, its solutions display an amazing variety of complex behaviour and turbulence, such as inverse energy cascades (from small to large scales), direct enstrophy cascades (from large to small scales) and zonostrophy cascades (with non-isotropic transfers). On the other hand, the model is simple enough to accommodate convenient mathematical features such as a non-canonical Hamiltonian structure \cite{We83,Ta09}, and the equation conserves total energy and enstrophy. More importantly for our study of resonances, the equation admits an infinite number of so-called ``travelling wave'' solutions. These are simply solutions of the linearised version of (\ref{eq:CHM}) in the form of a travelling wave (called Rossby wave), parameterised by wavevectors $(k,l) \in \mathbb{R}^2:$
\begin{eqnarray}
\label{eq:basis}
\Psi_{(k,l)}(x,y,t) &=& e^{i \left(k\, x + l \,y - \omega(k,l)\, t \right)}\,,\\
\label{eq:dispersion}
 \omega(k,l)  & \equiv & -\frac{\beta \, k}{k^2+l^2}\,.
\end{eqnarray}
It is easy to check that $\Psi_{(k,l)},$ as well as its real part, satisfies also the full equation (\ref{eq:CHM}), i.e., the nonlinear term in that equation is identically zero. However, an arbitrary linear combination of travelling wave solutions with wavevectors that are not collinear, is still a solution of the linearised version of (\ref{eq:CHM}) but is not anymore a solution of the full equation.

The last observation leads to the study of the nonlinear term and the mixing of modes, and is the starting point in the construction of so-called resonant triad solutions, which are approximate solutions, valid in the asymptotic limit when the oscillation amplitudes are small. Without going into the details of the multiple-scales method \cite{Na08}, these solutions are linear combinations of three travelling waves, of the form:
\begin{equation}
\Phi(x,y,t) = \Re \left(A_1(t) \Psi_{(k_1,l_1)}(x,y,t) + A_2(t) \Psi_{(k_2,l_2)}(x,y,t) + A_3(t) \Psi_{(k_3,l_3)}(x,y,t)\right)\,,
\end{equation}
where $\Re$ denotes real part, $A_j$ are some complex functions of time that are ``slow'' compared to the waves, and the set of wave vectors satisfy the following system of equations:
\begin{eqnarray}
\label{eq:reso_1} k_1 + k_2 &=& k_3\\
\label{eq:reso_2} l_1 + l_2 &=& l_3\\
\label{eq:reso_3} \omega_1 + \omega_2 &=& \omega_3\,,
\end{eqnarray}
 where $\omega_j \equiv \omega(k_j,l_j)\,,\quad j= 1, 2, 3\,.$ Any set of three wavevectors satisfying equations (\ref{eq:reso_1})--(\ref{eq:reso_3}) is called a resonant triad. We remark that there is a consistency condition that leads to a nonlinear system of evolution equations for the three slow functions $A_j.$ We refer the reader to the vast literature on the subject, such as \cite{Na08, craik1988wave} and references therein. In a more general setting, these slow functions also depend on the space coordinates \cite{Ne69}, again `more slowly' than the travelling waves.

It is possible to extend the resonant triad solutions to a more general approximate solution, consisting of a combination of groups of resonant triads:
 \begin{equation}
 \label{eq:soluCluster}
 \Phi(x,y,t) = \Re \sum_{I=1}^N A_I(t) \Psi_{(k_I,l_I)}(x,y,t)\,,
 \end{equation}
 in such a way that any wavevector $(k_I,l_I)$ appearing in the combination above, belongs to at least one resonant triad. In this setting, several resonant triads may be connected, forming so-called clusters. The main question of this paper is the classification of all possible values of wavevectors $(k_I,l_I)$ that belong to resonant and near-resonant triads when the spatial domain is periodic. However, it is worth mentioning that the $N$ functions $A_I(t)$ satisfy a coupled nonlinear system of evolution equations, that is derived using a set of consistency conditions, in a straightforward manner \cite{Ka07, Bu09}.

\subsection{Periodic Domains and Numerical Simulations}
\label{subsec:period}

Periodic spatial domains are the common place for numerical simulations. One of the most reliable existing methods for periodic domains is the pseudo-spectral method, that takes advantage of the fast Fourier transform to compute partial derivatives and nonlinear terms. For quadratic nonlinearities, as in the CHM equation, conservation of energy as well as enstrophy is guaranteed. The pseudo-spectral method is widely used in numerical experiments due to its rapid convergence in terms of accuracy, and is preferred over finite-difference and finite-volume methods for fundamental studies, such as turbulence and weak wave turbulence.

In the schematic version of the pseudo-spectral method, solutions to equation (\ref{eq:CHM}) are sought on the periodic domain $(x,y) \in [0,2 \pi)\times[0,2 \pi)$, so solutions $\psi(x,y,t)$ satisfy $\psi(x,y,t) = \psi(x+2 \pi,y, t) = \psi(x,y+2 \pi,t).$  The numerical method begins by approximating the solution $\psi(x,y,t)$ as a finite sum of Fourier modes:
\begin{equation}
 \label{eq:Fourier}
\psi(x,y,t) = \Re \sum_{k=0}^{N_x} \sum_{l=-N_y}^{N_y} A_{(k,l)}(t) \Psi_{(k,l)}(x,y,t),
\end{equation}
where a ``mode'' $\Psi_{(k,l)}(x,y,t)$ is defined in equation (\ref{eq:basis}) and is identified with its associated wavevector $(k,l).$ The unknown coefficients $A_{(k,l)}(t)$ need to be advanced in time numerically, using high-order techniques such as $4^{th}$-order Runge-Kutta. The natural numbers $N_x, N_y$ denote the spatial resolution of the numerical approximation, so that oscillations of wavenumber greater than these numbers cannot be resolved by the numerical scheme.

\section{Discrete Resonant Triads}
\label{sec:discrete}

In the periodic setting discussed in Section \ref{subsec:period}, the relevant wavevectors $(k,l)$ belong to a discrete lattice of integer numbers, so the resonant conditions (\ref{eq:reso_1})--(\ref{eq:reso_3}) are now a system of Diophantine equations. For Rossby waves, the dispersion relation (\ref{eq:dispersion}) applies, and after some algebra the resonant conditions can be reduced to the following equation:
\begin{equation}
\label{eq:simplif}
 k_3 (k_1^2+l_1^2)^2 - 2 k_3 (k_1^2+l_1^2)(k_3\,k_1 + l_3\,l_1) + 2 k_1 (k_3\,k_1 + l_3\,l_1)(k_3^2+l_3^2) - k_1 (k_3^2+l_3^2)^2 = 0\,,
\end{equation}
which can be interpreted as an equation for the pair $(k_1,l_1),$ if the pair $(k_3,l_3)$ is given. Here, the pair $(k_2,l_2)$ is obtained a posteriori by writing $(k_2,l_2) = (k_3,l_3)-(k_1,l_1).$\\

\noindent \textbf{Discarding Zonal Modes.} A mode $(k,l)$ with $k=0$ represents a so-called zonal mode. When $k_3=0,$ equation \eref{eq:simplif} reduces to $(2 l_1 - l_3)k_1  = 0,$ which is easy to solve. This case is not considered here because resonant interactions with zonal modes are trivial: physically these interactions are suppressed due to the vanishing of interaction coefficients, see for example \cite[equation 13.9]{Naz}. We remark that zonal flows are interesting, though, if quasi-resonant interactions are allowed (see Section \ref{sec:quasi}).\\

So, from here on we will assume $k_1,k_2,k_3 \neq 0.$ Without loss of generality one can assume also that $0 < k_1 \leq k_2 \leq k_3,$ but this latter assumption is not essential until the counting of physically different solutions of equation (\ref{eq:simplif}) is performed. \\

\noindent \textbf{Irreducible Triads.} Notice that equations (\ref{eq:simplif}) are invariant under overall re-scaling of the wavenumbers. This is useful from the computational point of view, because it means that we need to search for so-called irreducible triads, which are defined as solutions of the resonant condition \eref{eq:simplif} such that the six wavenumbers $k_1, l_1, k_2,l_2,k_3,l_3$ do not have a common factor.
Correspondingly, a reducible triad is defined by any solution that is an integer multiple of an irreducible triad.\\

\noindent \textbf{Change of Basis.} Let us make a convenient change of basis to the orthogonal basis spanned by $(k_3,l_3)$ and $(-l_3,k_3)$:
\begin{equation}
\label{eq:transf}
 (k_1,l_1) = a \,(k_3,l_3) + b\,(-l_3,k_3),
\end{equation}
where $(a,b) \in \mathbb{Q}^2\,.$ Explicitly, the inverse transformation is
\begin{equation}
\label{eq:transf_inv}
a = \frac{k_3\,k_1 + l_3\,l_1}{k_3^2+l_3^2}\,, \qquad b = \frac{k_3\,l_1 - l_3\,k_1}{k_3^2+l_3^2}.
\end{equation}
Notice that $k_1^2+l_1^2 = (a^2+b^2)(k_3^2+l_3^2)$ and $k_1 = a\,k_3 - b\,l_3\,.$ Therefore, equation (\ref{eq:simplif}) transforms to:
\begin{equation}
\label{eq:ab}
(a^2+b^2)^2 - 2\,a\,(a^2+b^2) + (2\,a-1)\left(a-\frac{l_3}{k_3}\,b\right) = 0\,.
\end{equation}
This equation is then equivalent to the original resonant condition, provided zonal modes are not involved.

\subsection{Mapping the resonance condition (\ref{eq:ab}) to classical problems: pure-cube solutions and elliptic curves}

Two cases need to be discussed separately:\\

\noindent \textbf{Case $a=0.$} Equation \eref{eq:ab} simplifies to
\begin{equation}
 b^4 + \frac{l_3}{k_3}\,b = 0\,.
\end{equation}
We discard the solution $b=0$ because it represents a triad that is formed by collinear modes, and in this case the interaction coefficients are identically zero (i.e., the mixed contribution stemming from the nonlinear terms in equation (\ref{eq:CHM}) vanish identically). So we take, without loss of  generality, $b \neq 0.$ We get then $b^3 = - \frac{l_3}{k_3}.$ But $b$ is a rational number, so we deduce that the case $a=0$ gives a nontrivial triad if and only if the ratio $\frac{l_3}{k_3}$ is a pure-cube rational. In such instance we obtain a triad of the form $(k_1,l_1), (k_2,l_2), (k_3,l_3),$ where
\begin{equation}
\label{eq:cube_soln}
 (k_1,l_1) = -\left(\frac{l_3}{k_3}\right)^{1/3}\,(-l_3,k_3), \quad (k_2,l_2) = (k_3,l_3)-(k_1,l_1),
\end{equation}
where an overall scaling factor might be needed in order that all wavenumbers be integer.\\

\noindent \textbf{Example:} $l_3=1, k_3=8.$ Then $b = -\left(\frac{l_3}{k_3}\right)^{1/3} = -\frac{1}{2}$ and so, applying equation \eref{eq:cube_soln} directly we get a preliminary triad:
$$(k_1,l_1)' = \left(\frac{1}{2},-4\right), \quad (k_2,l_2)' = \left(\frac{15}{2},5\right)\,, \quad (k_3,l_3)' = (8,1)\,,$$
so we multiply by the factor $2$ in order to get an integer irreducible triad:
$$(k_1,l_1) = \left(1,-8\right), \quad (k_2,l_2) = \left(15,10\right)\,, \quad (k_3,l_3) = (16,2)\,.$$

This ``pure-cube'' construction can be done for arbitrarily large values of wavenumbers and it leads to an infinite set of different irreducible resonant triads. But most of the solutions of the resonant conditions \eref{eq:ab} are found via a completely different method, to be discussed in the remaining of this Section.\\

\noindent \textbf{Case $a\neq 0.$} Provided the ratio $l_3/k_3$ is not a pure-cube rational, we define $\xi \equiv \frac{b}{a}$ and thus equation \eref{eq:ab} becomes
\begin{equation}
 a^3(1+\xi^2)^2 - 2\,a^2\,(1+\xi^2) + (2\,a-1)\left(1 -\frac{l_3}{k_3}\,\xi\right) = 0\,.
\end{equation}
Notice that this equation has degree 3 in $a$, one less than the original equation. We follow a standard procedure to isolate the coefficient of the cubic term. Defining $r \equiv a (1+\xi^2)$ we get
\begin{equation}
 r^3 - 2\,r^2 + (2\,r-(1+\xi^2))\left(1 -\frac{l_3}{k_3}\,\xi\right) = 0\,.
\end{equation}
Defining now $D \equiv \frac{1}{1 -\frac{l_3}{k_3}\,\xi}$ and the quantities
$X \equiv r D, \quad Y \equiv \xi D\,,$ we obtain the equation
\begin{equation}
\label{eq:ellicur}
 X^3 - 2\,D\,X^2 + 2\,D\,X - D^2 = Y^2\,,
\end{equation}
which can be interpreted as an Elliptic Curve if $D$ is fixed and we let $X, Y$ be variable.\\

\noindent \textbf{Summary of case $a\neq 0.$} To summarize, provided the ratio $l_3/k_3$ is not a pure-cube rational, there is a rational mapping from an integer triad $(k_1,l_1), (k_2,l_2), (k_3,l_3)$ satisfying all resonance conditions (\ref{eq:reso_1})--(\ref{eq:reso_3}), to the variables of the elliptic curve \eref{eq:ellicur}. The mapping is bijective up to overall re-scaling of the triad wavenumbers, and given explicitly by:
\begin{equation}
 \label{eq:map}
X = \frac{k_3}{k_1} \times \frac{k_1^2+l_1^2}{k_3^2+l_3^2}\,, \quad Y = \frac{k_3}{k_1} \times \frac{k_3 l_1 - k_1 l_3}{k_3^2+l_3^2}\,, \quad D = \frac{k_3}{k_1} \times \frac{k_3 k_1 + l_3 l_1}{k_3^2+l_3^2},
\end{equation}
and with inverse
\begin{equation}
 \label{eq:map_inv}
\frac{k_1}{k_3} = \frac{X}{D^2+Y^2}\,,\quad \frac{l_1}{k_3} = \frac{X}{Y} \left(1-\frac{D}{D^2+Y^2}\right)\,,\quad \frac{l_3}{k_3} = \frac{D-1}{Y}\,,
\end{equation}
and $(k_2,l_2) = (k_3-k_1,l_3-l_1).$

\subsection{Classification of solutions of triad equations in terms of Fermat's theorem of sums of squares}
\label{sec:Fermat}

We consider the case $a \neq 0$ described in the previous Section. We can rewrite the elliptic curve \eref{eq:ellicur} as:
\begin{equation}
 \label{eq:ellicur1}
Y^2 + \left(D + X^2 - X\right)^2 = X^2 \left(X^2-X+1\right)\,,
\end{equation}
and we can divide by $X^2$ because of our assumption of no zonal modes. We obtain:
\begin{equation}
 \label{eq:ellicur2}
\left(\frac{Y}{X}\right)^2 + \left(\frac{D}{X} + X - 1\right)^2 = X^2-X+1\,.
\end{equation}
The LHS of equation \eref{eq:ellicur2} is a sum of squares of rationals. The RHS is a quadratic form that is best written in diagonal form by defining
\begin{equation}
\label{eq:Xmn}
X \equiv -\frac{m+n}{m-n}, \quad m, n \in \mathbb{Z}\,,
\end{equation}
so we get the equation:
\begin{equation}
 \label{eq:ellicur3}
\left(\frac{Y (m-n)^2}{m+n}\right)^2 + \left(\frac{D(m-n)^2}{m+n} + 2\,m\right)^2 = 3\,m^2 + n^2\,.
\end{equation}
Equation \eref{eq:ellicur3} is to be solved for $m,n, \in \mathbb{Z}$ and $Y, D \in \mathbb{Q}.$ The cases $Y=0,$ $m=0$ or $m= \pm n$ are excluded from the solutions: they give rise to zonal modes. Notice that since $m, n$ are integers, the two members of equation \eref{eq:ellicur3} are equal to an integer. The problem of finding all possible integers that can be written as a sum of squares of two rationals is well known and dates back to Fermat \cite[Chapter V]{Dickson} (it is called Fermat's Xmas theorem). Also, the problem of finding all possible integers in the form $3\,m^2 + n^2$ was considered by Fermat and solved by Lagrange and Euler, see \cite[Chapter 1]{Cox} for further details. The current equation is a combination of these two problems and can be dealt with in a straightforward manner. As a result, all possible representations for the numbers $m, n$ and the numbers $Y, D$ can be obtained explicitly. Consequently, all possible exact resonant triads can be obtained explicitly by using the mapping \eref{eq:map_inv} from $X, Y ,D$ to the triad's wavenumbers. \\

\noindent \textbf{Method for finding the general solution of equation \eref{eq:ellicur3}.} We believe that it is not illuminating to provide here a detailed exposition of the solution method. We have relegated to the Appendix all necessary Theorems and Corollaries. Here we provide only the main idea of the solution method, but we still give the detailed explicit construction of the solution. The main idea is that integers of the form \eref{eq:ellicur3}, LHS, must be products of prime numbers of the form $4 K+1$ with $K$ some integer, times a square. On the other hand, integers of the form \eref{eq:ellicur3}, RHS, must be products of prime numbers of the form $3 K'+1$ with $K'$ some integer, times another square. So equating the two members of equation \eref{eq:ellicur3} we deduce that these must be equal to an integer that is product of primes of the form $12 K'' + 1$ with $K''$ some integer, times a square.

It follows from Corollary 4 in the Appendix, that all possible solutions of equation \eref{eq:ellicur3} for $m,n \in \mathbb{Z}$ coprime and $Y, D \in \mathbb{Q},$ can be parameterised explicitly using the following algorithm:

\begin{enumerate}
 \item  Construct the following expansion in prime powers:
\begin{equation}
 \label{eq:Ngen}
 N = 4^{a_0}\times{ \prod\limits_{j=1}^{M}} p_j^{a_j} \times\left({ \prod\limits_{j=1}^{M'}} q_j^{b_j}\right)^2 \times\left({ \prod\limits_{j=1}^{M''}} r_j^{c_j}\right)^2\,,
\end{equation}
where:

\noindent $a_0$ can take the values  $0$ or $1;$

\noindent  $\{p_j\}_{j=1}^\infty$ is the set of primes of the form $12\,k+1$ with $k$ integer;

\noindent  $\{q_j\}_{j=1}^\infty$ is the  set of primes of the form $3\,k+1$ with $k$ integer, excluding the primes $p_j$;

\noindent  $\{r_j\}_{j=1}^\infty$ is the set of primes of the form $4\,k+1$ with $k$ integer, excluding the primes $p_j$;

\noindent  The non-negative integers $M, M', M''$ denote how many primes are there in the expansion of $N$: the explicit expansion is given via three strings of non-negative integers $\{a_j\}_{j=1}^M$, $\{b_j\}_{j=1}^{M'}$ and $\{c_j\}_{j=1}^{M''}.$ These are defined so that if $M=0$ then the string $\{a_j\}_{j=1}^M$ is empty, and if $M > 0$ then $a_M > 0.$ Similar relations hold for $M$ replaced by $M'$ (resp. $M''$) and $a_j$ replaced by $b_j$ (resp. $c_j$).

Once $a_0$ and the strings $\{a_j\}_{j=1}^M$, $\{b_j\}_{j=1}^{M'}$ and $\{c_j\}_{j=1}^{M''}$ are known, the corresponding $N$ is uniquely constructed.

\item Construct all possible solutions $m,n \in \mathbb{Z}$ of the equation
\begin{equation}
 \label{eq:NmnAlg}
 N = 3\,m^2 + n^2\,,
\end{equation}
using the Brahmagupta identity \eref{eq:Brahma} on the individual expansions of the solutions of $p_j = 3 \,m_j^2 + n_j^2$ and $4 = 3 \times 1^2 + 1^2.$

The choice of sign for $m$ leads to two sets of physically sensible solutions, and one can take $n > 0$ without loss of generality. With this taken into account, it can be shown that the number of different solutions for $m,n$ (with $m \neq 0, \pm n$) is equal to
$$2\times \left(2 a_0 + 1\right) \times \frac{1}{2}\times \left[\prod_{j=1}^M \prod_{k=1}^{M'} \left(a_j + 1\right)  \left(2\,b_k + 1\right) - \epsilon\right],$$
 where   $\epsilon = 1$ if all $\{a_j\}_{j=1}^M$ are even, and $\epsilon = 0$ otherwise.

\item Construct all possible solutions $S,Q \in \mathbb{Z}$ of the equation
\begin{equation}
 \label{eq:NSQAlg}
 N = S^2 + Q^2\,,
\end{equation}
using the Brahmagupta identity \eref{eq:Brahma} on the individual expansions of the solutions of $p_j = S_j^2 + Q_j^2.$

We can take without loss of generality the convention $0 \leq S < Q.$ With this taken into account, it can be shown that the number of different solutions for $S,Q$ is equal to
$$\frac{1}{2}\times \left[\prod_{j=1}^M \prod_{l=1}^{M''} \left(a_j + 1\right)\left(2\,c_l + 1\right) - \epsilon\right] + \epsilon,$$
where $\epsilon$ was defined in point (ii). Notice that the factor $4^{a_0}$ does not play an important role here.

\item Based on the above solutions for $S,Q$ integers, construct all possible solutions $s,q \in \mathbb{Q} \setminus \mathbb{Z}$ of the equation
\begin{equation}
 \label{eq:NsqAlg}
 N = s^2 + q^2\,,
\end{equation}
using the Brahmagupta identity \eref{eq:Brahma} on the solutions obtained in point (iii) along with the Pythagorean rationals, which are defined by the remarkable identity:
\begin{equation}
 \label{eq:elegant1}
1 = \left(\frac{2 \,u\,v}{u^2+v^2}\right)^2 + \left(\frac{u^2 - v^2}{u^2+v^2}\right)^2\,, \quad \forall \,u, v \in \mathbb{Z}.
\end{equation}
 The relevant solutions for $s,q$ are of the form:
\begin{equation}
 \label{eq:solNsqAlg}
 s = \frac{2\,u\,v \,Q + (u^2-v^2)\,S}{u^2+v^2}\,,\quad  q = \frac{- 2\,u\,v \,S + (u^2-v^2)\,Q}{u^2+v^2}\,,
\end{equation}
where $u,v$ are coprime integers satisfying $1 \leq |v| < u.$ This ordering prevents repetition of solutions, which at this stage we need to avoid, although they play a role in point (v).

For computational purposes we parameterise the coprime integers $u,v$ using finite sets of adjustable size. Letting $U_{\max} \in \mathbb{Z},$ $U_{\max} \geq 2,$ define the ``Pythagorean Fan'' $F(U_{\max})$ by the equation
$$F(U_{\max}) \equiv \{ (u,v) \in \mathbb{Z}^2 \quad \mathrm{coprime} \quad \mathrm{s.t.} \quad 1 \leq |v| < u \leq U_{\max}\}\,.$$

Then, for a given solution $S,Q$ of equation \eref{eq:NSQAlg}, a set of non-integer rational solutions $s,q$ of equation \eref{eq:NsqAlg} is obtained which amounts to $\# F(U_{\max})$ new solutions, where $\# F(U_{\max})$ is the cardinality of $F(U_{\max}).$

\item Gather all results above and construct all possible solutions for $Y$ and $D$ of equation \eref{eq:ellicur3}. Here, an extra $8$-fold symmetry is available that allows us to generate more solutions and correspondingly more triads. Notice that for each obtained solution $s,q$ (integer as well as non-integer) of equation \eref{eq:NsqAlg}, and solution $m,n$ of equation \eref{eq:NmnAlg}, the following statement is true:  {Provided $s q \neq 0$}, we have the choice of writing eight different equations for $Y,D,$ schematically written as:
\begin{eqnarray}
 \label{eq:SolYD1}
\frac{Y (m-n)^2}{m+n} = \pm s\,,\quad \frac{D(m-n)^2}{m+n} + 2\,m = \pm q\,,
\end{eqnarray}
and
\begin{eqnarray}
 \label{eq:SolYD2}
\frac{Y (m-n)^2}{m+n} = \pm q\,,\quad \frac{D(m-n)^2}{m+n} + 2\,m = \pm s\,,
\end{eqnarray}
where all sign combinations can be taken. Notice that the symmetry $Y \to - Y$ amounts to taking the mirror image of a triad about the $k$-axis in the $(k,l)$-wavevector space (see Eq.~\eref{eq:map_inv}).

There is a special case which occurs when all $a_j$'s are even. There, amongst all possible solutions $S,Q$ there is one with $S = 0.$ Moreover, even when $S Q \neq 0$ there is a solution $s,q$ with $s q = 0$ for some choice of $u,v.$ In these cases we still have a symmetry, but it reduces to a $2$-fold symmetry, with two different equations for $Y,D$:
\begin{eqnarray}
 \label{eq:SolYD3}
\frac{Y (m-n)^2}{m+n} = \pm s\,,\qquad \frac{D(m-n)^2}{m+n} + 2\,m = 0\,.
\end{eqnarray}
\end{enumerate}

Gathering all solutions together, we can write a formula for the total number of solutions for $X,Y,D$ of equation \eref{eq:ellicur1}, leading to the following number of irreducible triads as a function of the integer $N = 4^{a_0}\times{ \prod\limits_{j=1}^{M}} p_j^{a_j} \times\left({ \prod\limits_{k=1}^{M'}} q_k^{b_k}\right)^2 \times\left({ \prod\limits_{l=1}^{M''}} r_l^{c_l}\right)^2$:
\begin{eqnarray}
\nonumber
 T(N) &=&  \left(2 a_0 + 1\right) \times \left[\prod_{j=1}^M \prod_{k=1}^{M'} \left(a_j + 1\right)  \left(2\,b_k + 1\right) - \epsilon\right] \\
 \label{eq:totalSols}
      &\times&  \left[\prod_{j=1}^M \prod_{l=1}^{M''} \left(a_j + 1\right)  \left(2\,c_l + 1\right) + \epsilon\right] \times (4\,\# F(U_{\max}) + 4 - 3 \, \epsilon).
\end{eqnarray}
For example, taking $N = 4 \times 13$ and $U_{\max} = 2,$ so $a_0=1, a_1 = 1, M = 1, M'=M''=0, \# F(U_{\max}) = 2, \epsilon = 0,$ we get $T(N) = 144$ triads.\\

\noindent \textbf{Remarks.}

\begin{itemize}
 \item As defined earlier, an irreducible triad is one whose wavenumbers $k_1, l_1, k_2, $ $l_2, k_3, l_3$ do not have a common factor. Knowing all irreducible triads within a given box in $(k,l)$-wavenumber space, say $|k|, |l| \leq L,$ allows one to compute all reducible triads within that box, by including integer multiples of irreducible triads of size smaller than $L/2$.

\item Nested character of the triads. Let the integers $N_1, N_2$ be of the form $4^{a_0}\times{ \prod\limits_{j=1}^{M}} p_j^{a_j} \times\left({ \prod\limits_{k=1}^{M'}} q_k^{b_k}\right)^2 \times\left({ \prod\limits_{l=1}^{M''}} r_l^{c_l}\right)^2.$ Then if $N_1 = N_2 \times N_3^2$ for some integer $N_3,$ then the set of triads obtained using $N_1$ contain the set of triads obtained using $N_2.$

\item Eliminating extra multiplicity. The set of all possible triads (using all possible values of $N$) obtained from this method, will appear ``repeated six times'' in the sense of the six-fold permutation symmetry of the triad equations:
\end{itemize}

\begin{eqnarray}
 \mathbf{k}_1 + \mathbf{k}_2 = \mathbf{k}_3 \,,&\quad & \mathbf{k}_2 + \mathbf{k}_1 = \mathbf{k}_3, \\
(-\mathbf{k}_3)+\mathbf{k}_2 = (-\mathbf{k}_1),& \quad &\mathbf{k}_2 + (-\mathbf{k}_3) = (-\mathbf{k}_1), \\
(-\mathbf{k}_3)+\mathbf{k}_1 = (-\mathbf{k}_2),& \quad &\mathbf{k}_1 + (-\mathbf{k}_3) = (-\mathbf{k}_2),
\end{eqnarray}
where $\mathbf{k}_j \equiv (k_j,l_j).$ The symmetry reduces to four-fold when  permutations of ``pure-cube'' triads occur, because a ``pure-cube'' triad cannot be mapped to an elliptic curve, but four of its permutations can be mapped. Correspondingly, when collecting all triads obtained using our method, a normalisation must be applied if one wants to use formula \eref{eq:totalSols} to estimate the number of genuinely different triads. The exact number of triads for a given $N$ is thus a bit different than what the formula predicts. Only when triads are collected using several values of $N,$ the formula \eref{eq:totalSols} will give asymptotically a good estimation if we divide by $6.$ As an example, the choice $N=4\times 13$ gives 40 genuinely different triads, rather than $144/6 = 24$.
%These triads are, explicitly, the 20 triads:
%$$\left(
%\begin{array}{ccc}
% \{1,8\} & \{15,-10\} & \{16,-2\} \\
% \{3,11\} & \{5,-25\} & \{8,-14\} \\
% \{5,25\} & \{27,-21\} & \{32,4\} \\
% \{3,41\} & \{117,-91\} & \{120,-50\} \\
% \{39,143\} & \{81,-93\} & \{120,50\} \\
% \{45,95\} & \{51,-187\} & \{96,-92\} \\
% \{1,47\} & \{135,-285\} & \{136,-238\} \\
% \{27,174\} & \{53,-424\} & \{80,-250\} \\
% \{1,83\} & \{159,-583\} & \{160,-500\} \\
% \{15,400\} & \{1377,-1734\} & \{1392,-1334\} \\
% \{45,1075\} & \{3699,-4247\} & \{3744,-3172\} \\
% \{1215,2310\} & \{1233,-4384\} & \{2448,-2074\} \\
% \{85,1525\} & \{4947,-4559\} & \{5032,-3034\} \\
%\{1649,5723\} & \{2295,-3345\} & \{3944,2378\} \\
% \{405,4605\} & \{843,-11521\} & \{1248,-6916\} \\
% \{2295,6405\} & \{3281,-13703\} & \{5576,-7298\} \\
% \{85,3025\} & \{9843,-16019\} & \{9928,-12994\} \\
% \{15,2200\} & \{2529,-16298\} & \{2544,-14098\} \\
% \{135,9465\} & \{289,-23987\} & \{424,-14522\} \\
 %\{5,4525\} & \{867,-34391\} & \{872,-29866\}
%\end{array}
%\right),$$
%along with their mirror images.

\section{Numerical results: Generating resonant triads with wavenumbers $(k,l)$ in the box $|k| \leq 5000, |l| \leq 5000$}
\label{sec:num_res}
Let us consider the following problem: Find all resonant triads in the box $|k|,|l| \leq 5000.$ As explained in the introduction, a direct search using a brute-force algorithm would take more than $15$ years to answer that problem. With the method introduced in this paper, such problem is reduced to the problem of \emph{finding the appropriate set of primes $p_j, q_j, r_j$ that generate all triads within the box.}

The answer to this latter problem has not been presented in this paper, and we believe this is a matter for a subsequent paper. From the rational character of the mapping from the triads to the points $X,Y,D$ on the elliptic curve, it follows that only small-enough primes $p_j,q_j,r_j$ and small-enough powers $a_j,b_j,c_j$ can generate small resonant triads. The open question is how small should the primes and the powers be. An extra complication is given by the Pythagorean Fan, which could in principle reduce the size of the triads.

Nevertheless, we believe that our new method can be used to obtain the vast majority of the triads within the given box, by collecting the triads obtained from combinations of small-enough primes $p_j,q_j,r_j$. We provide numerical support of this belief in the next paragraphs. Of course, we cannot wait the $15$ years in order to check explicitly the percentage of triads in the box $|k|,|l| \leq 5000$ that are generated with our new method. For this reason, to prove the point we performed a brute-force search to find all resonant triads in the box $|k|,|l| \leq 100.$ This took about $40$ minutes using a desktop computer. In contrast, on the same computer our method takes only $1/2$ of a second to generate the resonant triads within the box (along with many more triads outside the box), using a non-parallel computation of the triads generated by the following 16 numbers: $N= 4 \times p_j \times 7^2 \times 5^2$ and $N = 4 \times p_j^2 \times 7^2 \times 5^2,$ with $j=1, \ldots, 8,$ along with the Pythagorean triples generated by $u=2, v=\pm 1,$ which is used to generate extra rational solutions of the sum of squares problem.

Following the insight provided by the ``small-box'' case, we have developed a search method that looks at the triads in the box $|k|,|l| \leq 5000$ that are generated by:

\noindent The 1600 numbers $N = 4 \times p_j \times q_k^2 \times r_l^2,$ with $j=1, \ldots, 100, \quad k = 1, \ldots, 4 \quad$ and $l=1,\ldots,4,$ combined appropriately with the 42 Pythagorean triples generated by $1 \leq |v| < u \leq 8.$ Computing time: 287 seconds. Triads obtained: 510 irreducible triads within the box.

\noindent The 627 numbers $N= 4 \times p_j \times 7^2 \times 5^2$ with $j=101, \ldots, 727,$ combined appropriately with the 42 Pythagorean triples generated by $1 \leq |v| < u \leq 8.$  Computing time: 117 seconds. Triads obtained: 66 irreducible triads within the box.

\noindent The 1792 numbers $N = 4 \times p_j^2 \times q_k^2 \times r_l^2,$ with $j=1, \ldots, 7, \quad k = 1, \ldots, 16 \quad$ and $l=1,\ldots, 16,$ combined appropriately with  the 42 Pythagorean triples generated by $1 \leq |v| < u \leq 8.$ Computing time: 665 seconds. Triads obtained: 134 irreducible triads within the box.

\noindent The 120 numbers $N = 4 \times p_j p_k \times 7^2 \times 5^2,$ with $j = 1,2,3$ and $ k = j+1 ,\ldots, j + 40,$ combined appropriately with the 42 Pythagorean triples generated by $1 \leq |v| < u \leq 8.$ Computing time: 122 seconds. Triads obtained: 160 irreducible triads within the box.

\begin{figure}
\begin{center}
\includegraphics[height=35mm]{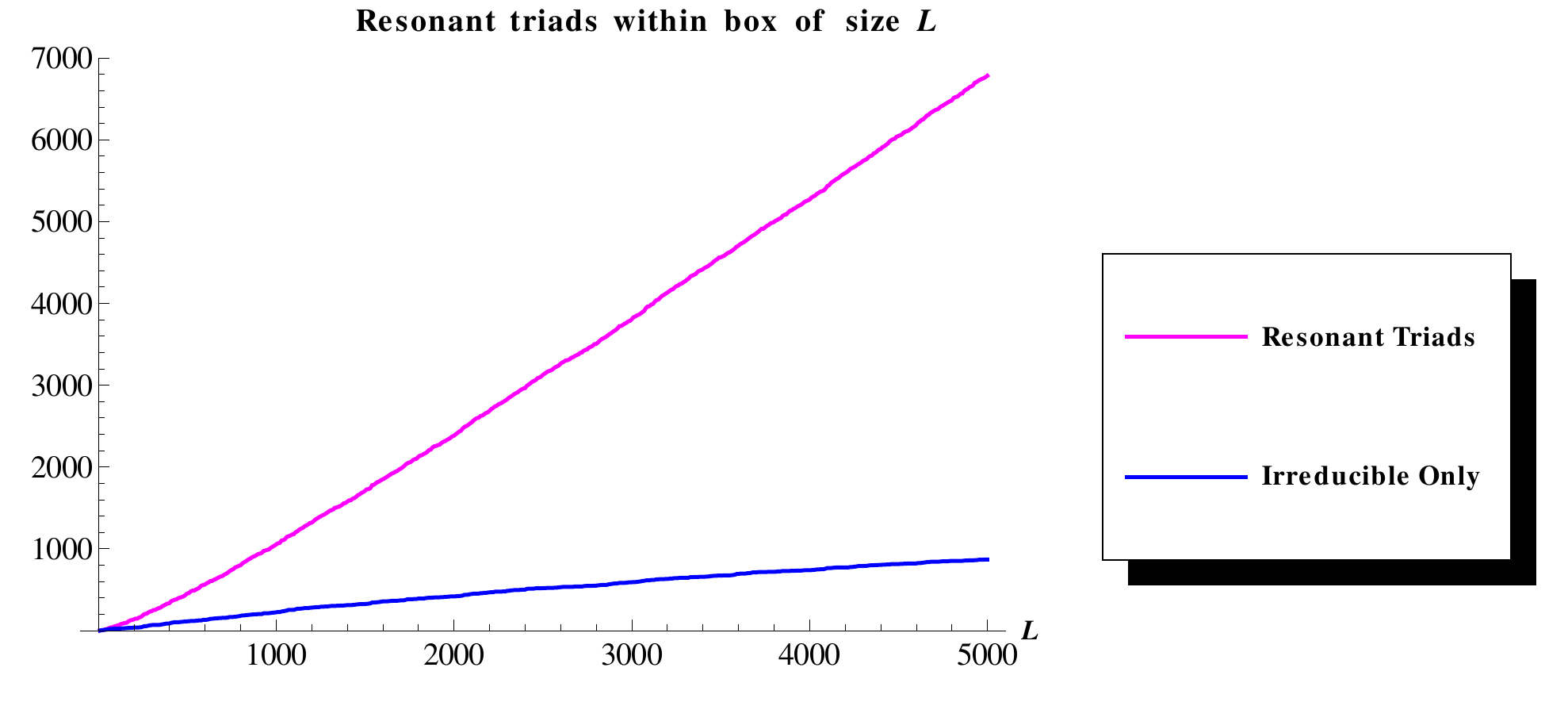}
\includegraphics[height=35mm]{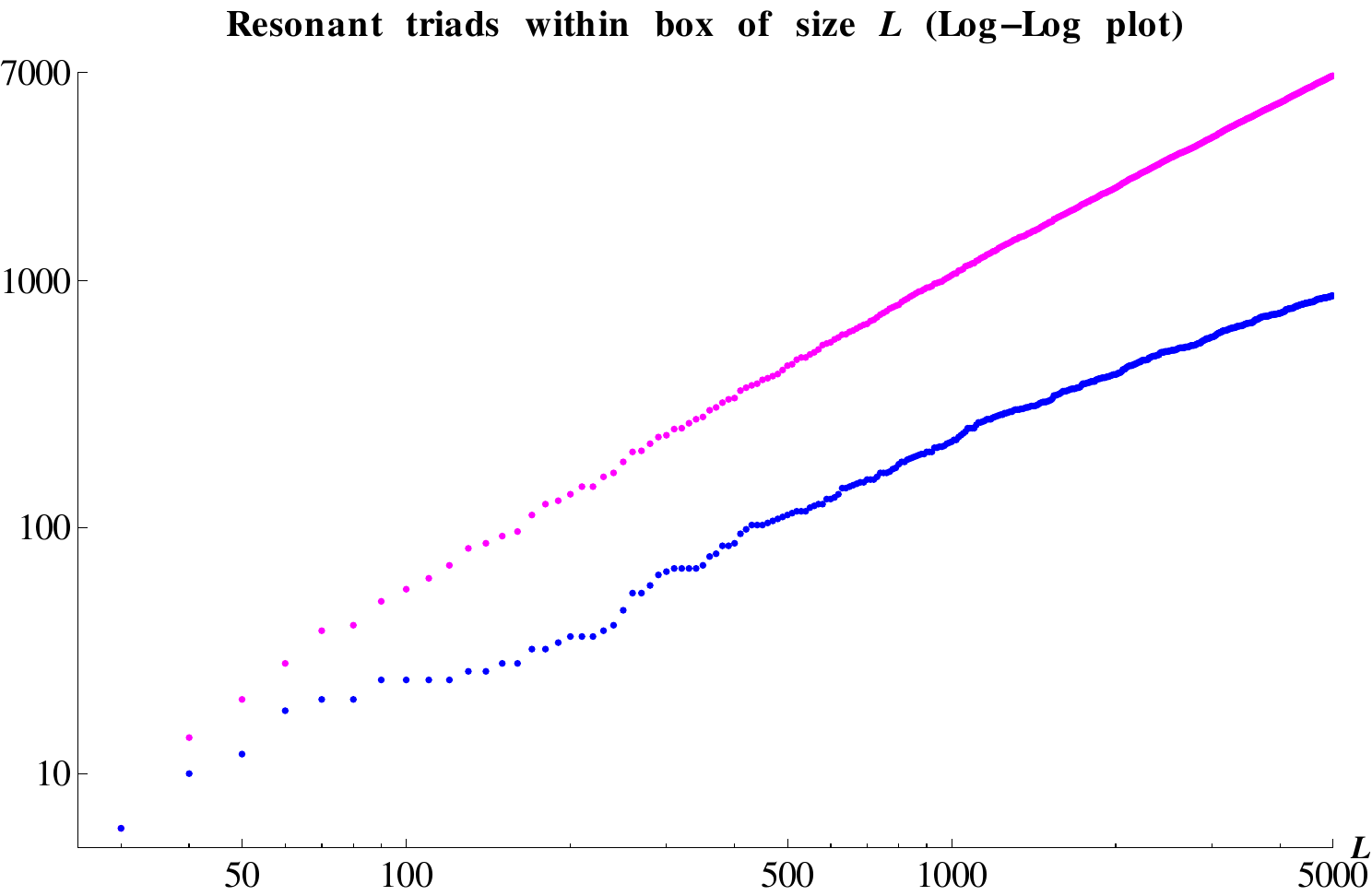}
\caption{\label{fig:number_triads} \textbf{Left panel:} Number of exact resonant triads within box of size $L$ (i.e., wavenumbers bounded by $|k|, |l| \leq L$), as a functon of $L.$ Black curve (blue online): irreducible triads only. Grey curve (magenta online): including reducible and irreducible triads. \textbf{Right Panel:} Same plot as left panel, but in log-log scale.}
\end{center}
\end{figure}

We have tested higher-degree combinations of primes and we have obtained up to 20 more triads, however for simplicity of presentation we do not consider these here. In total, our search takes about $20$ minutes on an $8$-core desktop computer using \emph{Mathematica}. We obtain a total of $870$ irreducible triads, which leads to a total of $6794$ triads (including reducible ones) in the box $|k|, |l| \leq 5000.$
A plot of (i) the number of irreducible resonant triads and (ii) the number of total resonant triads, contained in a box as a function of the box size, is shown in Figure \ref{fig:number_triads}, left panel. The right panel plots the same variables in log-log scaling.

It is apparent from figure \ref{fig:number_triads}, right panel, that the total number of resonant triads (including reducible ones) as a function of the box size, behaves as a power law of the size, in the form: $T = C L^\alpha$, where $\alpha \approx 1.2$ and $C \approx 0.263$ (this is obtained using fit interval $200 \leq L \leq 4000$). Notice that our numerical method gives an exact estimate for the number of triads for $L=100,$ and the estimation is by construction more accurate for smaller $L$ than for larger $L$. So, obtaining a power law for the number of triads over a range of values of $L$ is a good indication of the accuracy of our estimations.

\begin{figure}
\begin{center}
\includegraphics[width=140mm]{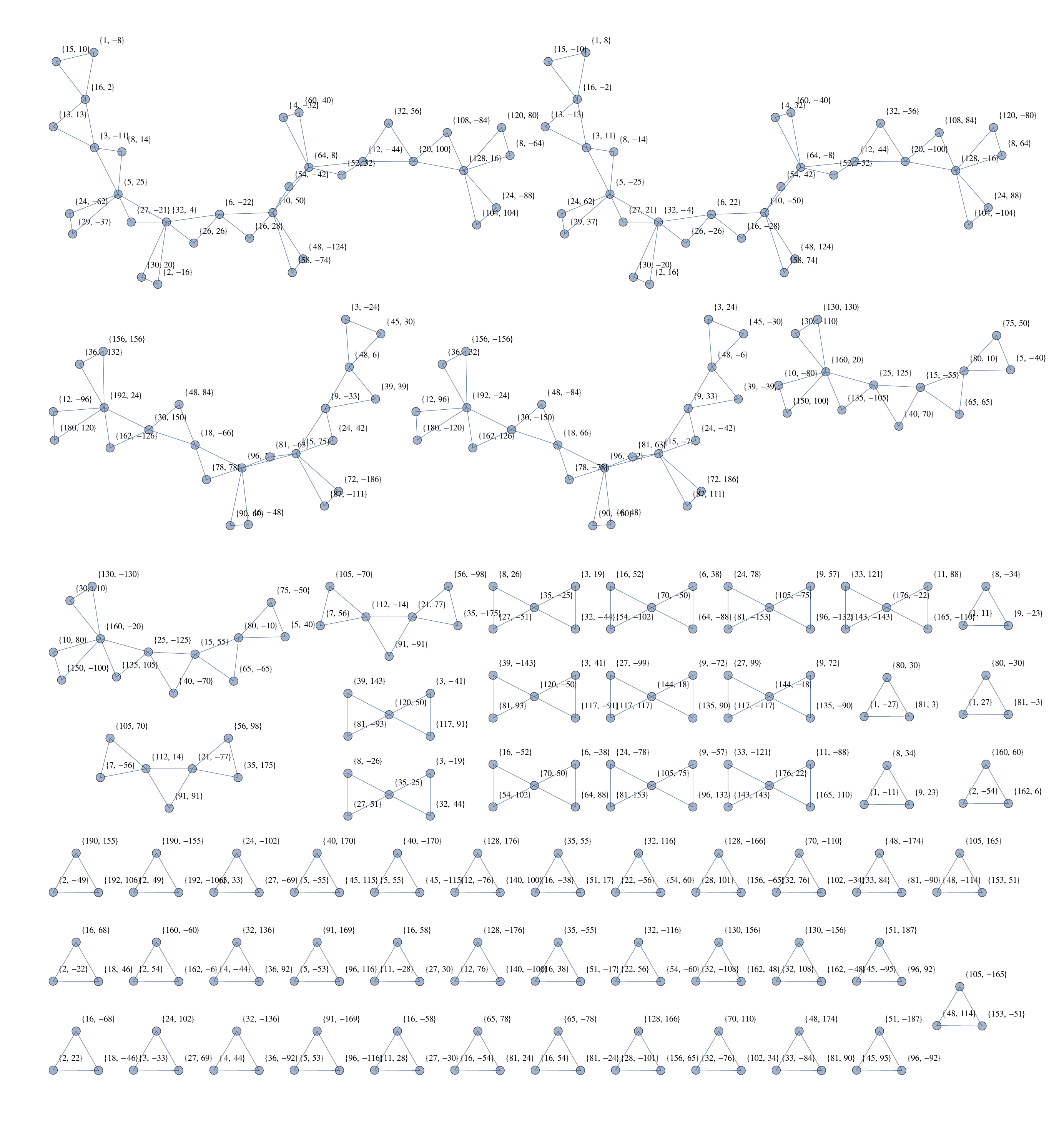}
\caption{\label{fig:reso_clusters_graph} Symbolic plot of the CHM resonant clusters generated within the box of size $L=200.$}
\end{center}
\end{figure}

\begin{figure}
\begin{center}
\includegraphics[height=40mm]{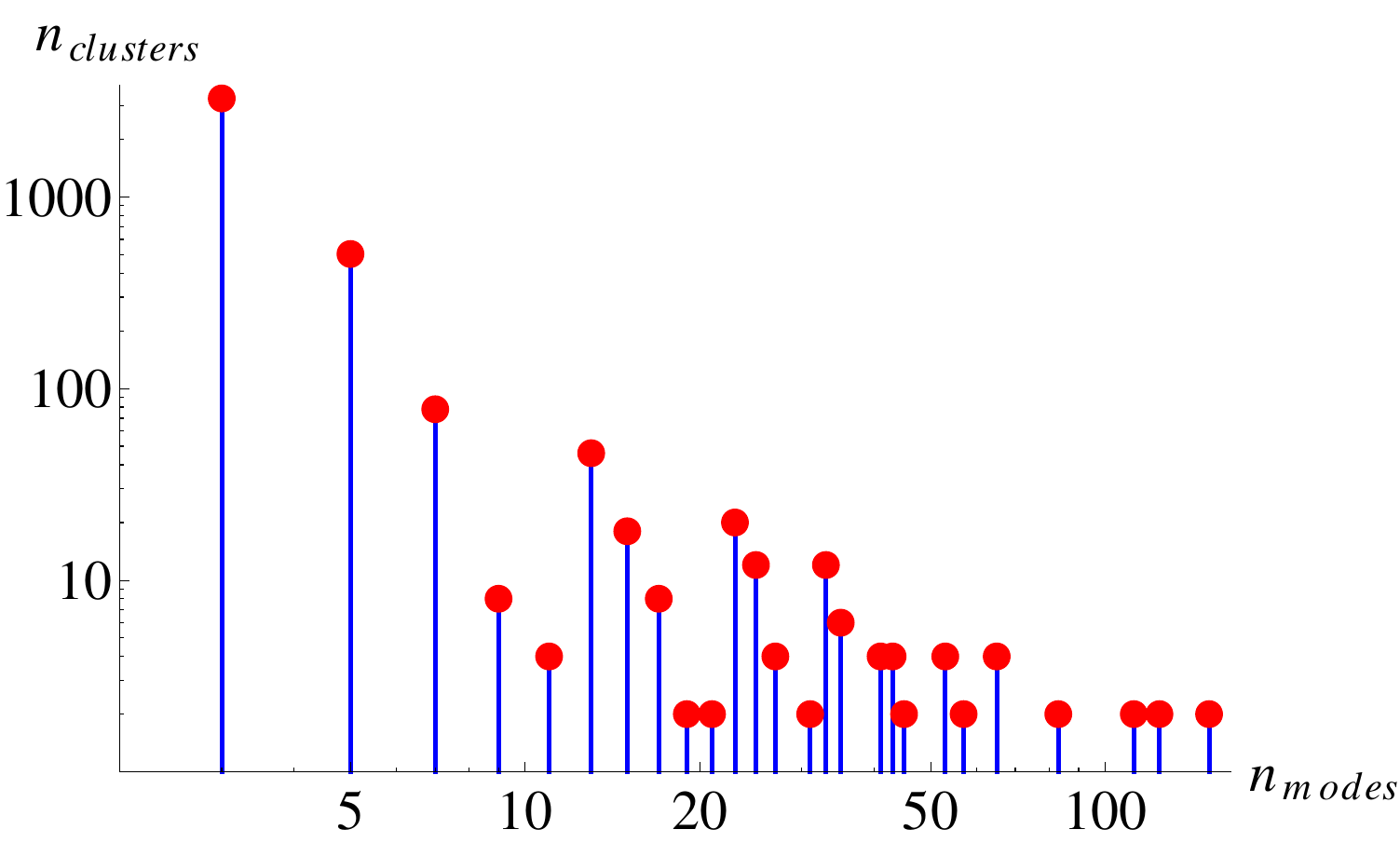}
\includegraphics[height=40mm]{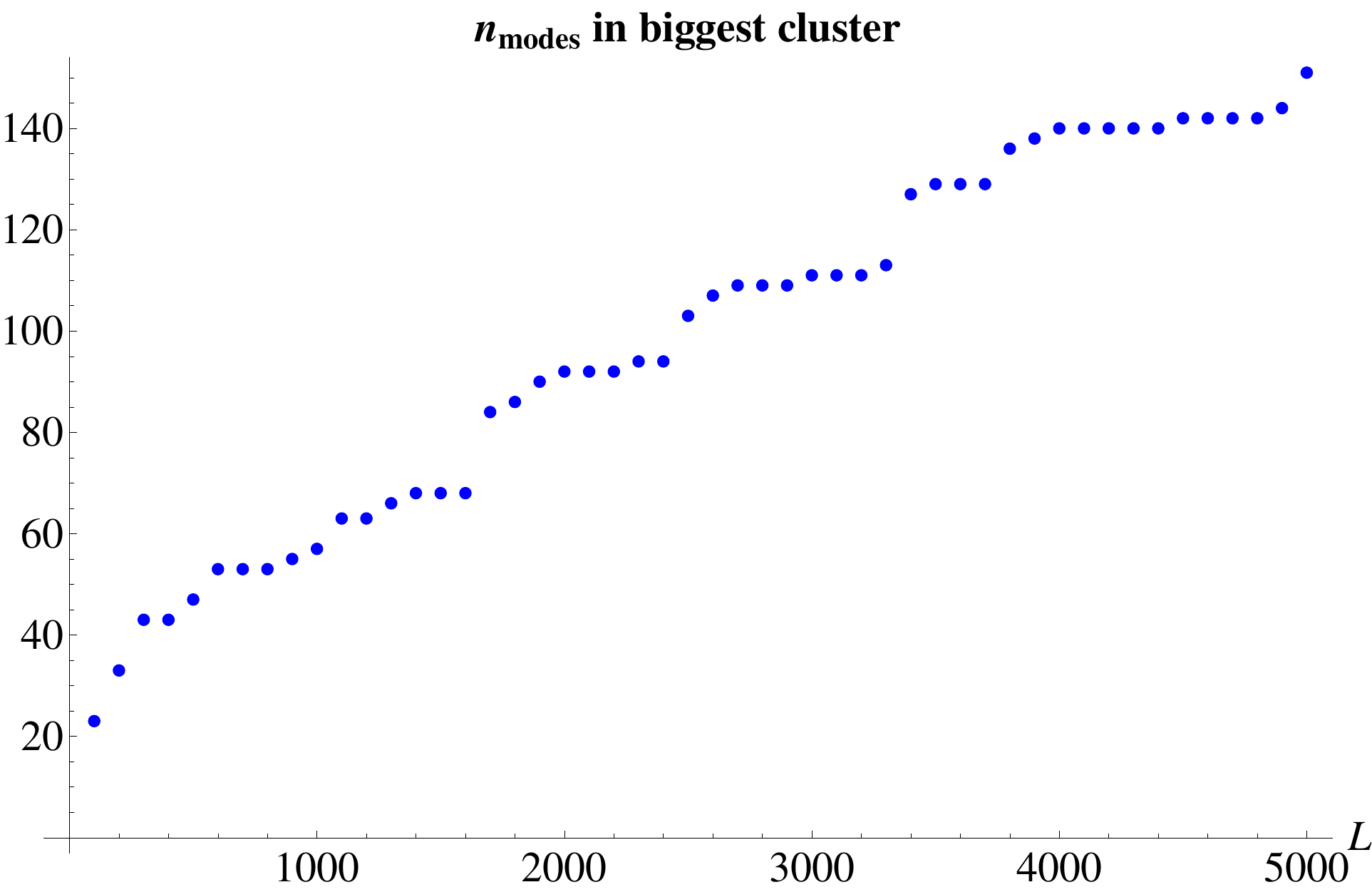}
\includegraphics[height=40mm]{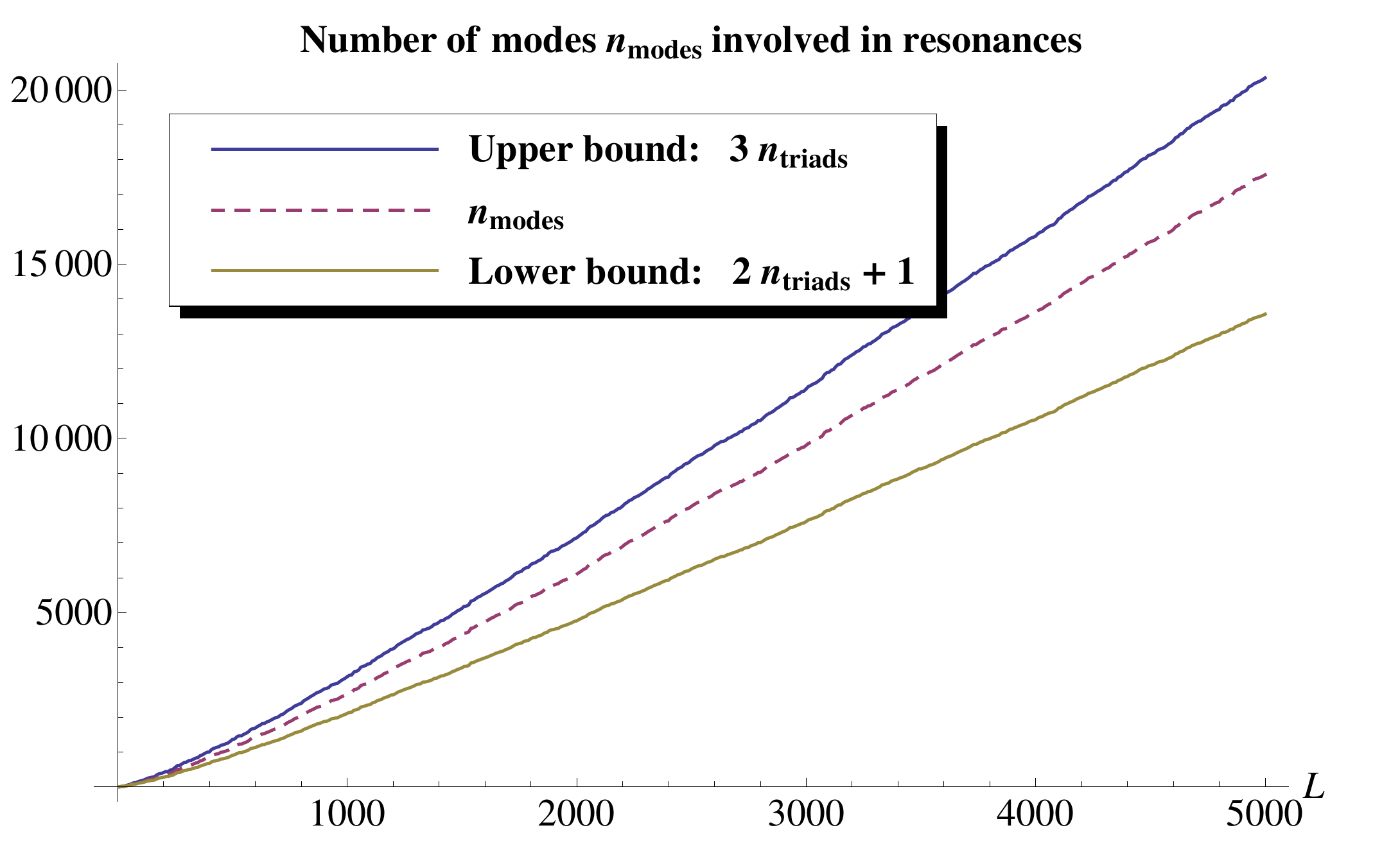}
\caption{\label{fig:reso_clusters} \textbf{Top Left Panel:} Circles (red online): Log-log plot of the distribution of number of clusters ($n_{\mathrm{clusters}}$-axis) as a function of cluster size ($n_{\mathrm{modes}}$-axis, the number of modes in the cluster), for the collection of resonant clusters found within the box of size $L=5000.$ For example, there are $8$ clusters formed by exactly $9$ modes each, and about $80$ clusters formed by exactly $7$ modes each. \textbf{Top Right Panel:} Number of modes in the largest resonant cluster found within the box of size $L$, as a function of $L.$ \textbf{Lower Panel:} Dashed curve (magenta online): number of modes $n_{\mathrm{modes}}$ involved in the collection of all resonant clusters found within the box of size $L$, as a function of $L.$ The upper and lower curves correspond to the ``sandwich'' bound $3\, n_{\mathrm{triads}} \geq n_{\mathrm{modes}} \geq 2 \,n_{\mathrm{triads}} + 1,$ valid when the connectivity of the triads in the clusters is at most of one-common-mode type.}
\end{center}
\end{figure}

Another aspect of the set of resonant triads within the box of size $L=5000$ is the connectivity of the web of resonant triads (including reducible ones). In figure \ref{fig:reso_clusters_graph}, we show a diagrammatic plot of the clusters that appear within a box of size $L=200$ (for higher values of $L$ the plots become prohibitively populated). We notice the predominance of isolated triads. This feature persists for larger box sizes, and the isolated triads contribute with approximately $50\%$ of the total number of resonant triads. However, there is a nontrivial distribution of cluster sizes within a given box size. This is evidenced in figure \ref{fig:reso_clusters}, top left panel, where we present the log-log plot of the distribution of number of clusters (y-axis) as a function of cluster size (i.e., the number of modes in the cluster), for the collection of resonant clusters within the box of size $L=5000.$ For example, isolated triads have a cluster size $n_{\mathrm{modes}}=3$ and butterflies have cluster size $n_{\mathrm{modes}}=5.$ It is apparent from the plot that several cluster sizes are represented, in a kind of power law. For reference, we plot the size of the biggest cluster as a function of box size $L$ in figure \ref{fig:reso_clusters}, top right panel. It is evident that the biggest cluster contributes with only a fraction of the modes involved in resonant interactions (about $1\%$). Finally, the total number of modes $n_{\mathrm{tot}}$ involved in resonant interactions within a given box of size $L$, as a function of $L$, behaves in a way that is similar to the total number of triads in the box. In fact, due to the fact that resonant triad connections occur via one shared mode at a time, it can be shown analytically that the total number of modes involved in resonant interactions is bounded between $2 n_{\mathrm{triads}} + 1 $ and $3  n_{\mathrm{triads}},$ where $n_{\mathrm{triads}}$ is the total number of triads. The origin of these bounds is as follows: the lower bound comes from the assumption that all modes are connected in a single cluster, \emph{with connections between triads via a single common mode}, as the clusters in figure \ref{fig:reso_clusters_graph}. The upper bound is the case when all modes form isolated triads. Finally, in figure \ref{fig:reso_clusters}, lower panel, we show the plot of the total number of modes involved in resonant interactions $n_{\mathrm{modes}}$ as a function of box size $L$, along with the respective bounds. Empirically, the relation between $n_{\mathrm{modes}}$ and $n_{\mathrm{triads}}$ is found to be linear (figure not shown). A linear fit (fitting interval $200 < L < 5000$) gives a relation $n_{\mathrm{modes}} \approx 2.6 \, n_{\mathrm{triads}} - 60.$

\section{Quasi-Resonant Triads}
\label{sec:quasi}
%Compare this empirical result with the rigorous result $T' \propto L^4,$ where $T'$ is the number of triads (resonant and non-resonant) in the box of size $L.$

Quasi-resonant triads are defined by the system of equations
\begin{eqnarray}
\label{eq:qreso_1} k_1 + k_2 &=& k_3\\
\label{eq:qreso_2} l_1 + l_2 &=& l_3\,,
\end{eqnarray}
along with the inequality
\begin{eqnarray}
\label{eq:qreso_3} |\omega_1 + \omega_2 - \omega_3| \leq \delta\,,
\end{eqnarray}
where the bound $\delta$ is known as the \emph{detuning level} allowed for a triad.

Quasi-resonant triads play a crucial role in the dynamics of CHM and are more important physically than the resonant triads. This is because in any realistic setting the dispersion relation has an experimental error, so in fact every physically sensible triad is quasi-resonant. Moreover, the amplitude of the wave oscillations is always finite, not infinitesimally small. Therefore the time scale of non-resonant triads is not infinitesimally small, and in fact it can be comparable with the time scale of the resonant nonlinear oscillations, provided $\delta$ is small enough. The accepted conclusion is that a full understanding of the dynamics of a wave system requires the understanding of the web of quasi-resonances, rather than just the web of resonances. Notice that the web of quasi-resonances for a given dispersion relation is robust (i.e. stable) under small perturbations of the dispersion relation, whereas the web of exact resonances is unstable under small perturbations.

These considerations would appear to imply that our method presented in this paper is useless. Fortunately, this is incorrect because our method can be used directly to generate quasi-resonant triads, in a hierarchical way in the sense that the triads generated have mismatch $\omega_1 + \omega_2 - \omega_3$ that is small.\\

\noindent \textbf{Numerical Method to generate quasi-resonant triads within a given box, starting from exact resonant triads of any size.} This is based on the fact that the dispersion relation in CHM, equation (\ref{eq:dispersion}) is homogeneous under overall re-scaling of the wavenumbers $k_1,l_1,k_2,l_2,k_3,l_3$ by any constant (not necessarily integer). Our previous method to generate exact resonant triads starts with a number $N$ and computes all possible representations of the prime expansion of this number $N$. Typically, one gets a lot of triads that are outside the given box. While these triads were discarded in the previous method, the new method uses all resonant triads available.

From here on we take $\beta=-1$ for simplicity in the dispersion relation (\ref{eq:dispersion}). Let $(K_1,L_1),(K_2,L_2), (K_3,L_3)$ be an irreducible resonant triad, with $K_j,L_j$ integers. Then the re-scaled triad $(\alpha K_1, \alpha L_1),(\alpha K_2,\alpha L_2), (\alpha K_3, \alpha L_3)$ is resonant, for any $\alpha \in \mathbb{R}.$ However this triad is not necessarily integer, so we need to approximate the scaled wavevectors to nearby integers, keeping in mind that equations (\ref{eq:reso_1}) and (\ref{eq:reso_2}) should be satisfied. This approximation will generate an error in the individual frequencies, so equation (\ref{eq:reso_3}) is not satisfied anymore but inequality (\ref{eq:qreso_3}) is satisfied, with detuning levels that depend on the accuracy of the approximation from re-scaled wavevectors to integer wavevectors.

In practice, suppose we want to generate quasi-resonant triads within a box of size $L$. We take an irreducible triad that is outside the box and re-scale it to just fit within the box (i.e., the maximum modulus of the six re-scaled components is now equal to $L$). The re-scaled components are written as real numbers in, say, double precision. This is our fundamental triad. Next, we take re-scaled copies of our fundamental triad, with box norms equal to $L-1, L-2, \ldots, 1.$ For each of these copies we approximate the triad to nearby integer triads, respecting the first two resonance conditions (\ref{eq:reso_1}) and (\ref{eq:reso_2}). The resulting set of triads is formed by quasi-resonant triads only, and the higher the box norm the smaller the size of the corresponding detuning level.

We repeat the process for every irreducible triad outside the box. In this way we can get a huge number of quasi-resonant triads (ordered by detuning level) starting from a comparatively small set of resonant triads.

The advantage of this new method is that it provides, in a matter of seconds, a set of quasi-resonant triads whose detuning levels are distributed smoothly about the origin. In contrast, the brute-force search for quasi-resonant triads takes a long time to generate a smooth distribution, of the order of the time it takes to scan all possible triads, resonant or non-resonant ($15$ years for a box of size $10000$).\\

\begin{figure}
\begin{center}
\includegraphics[width=70mm]{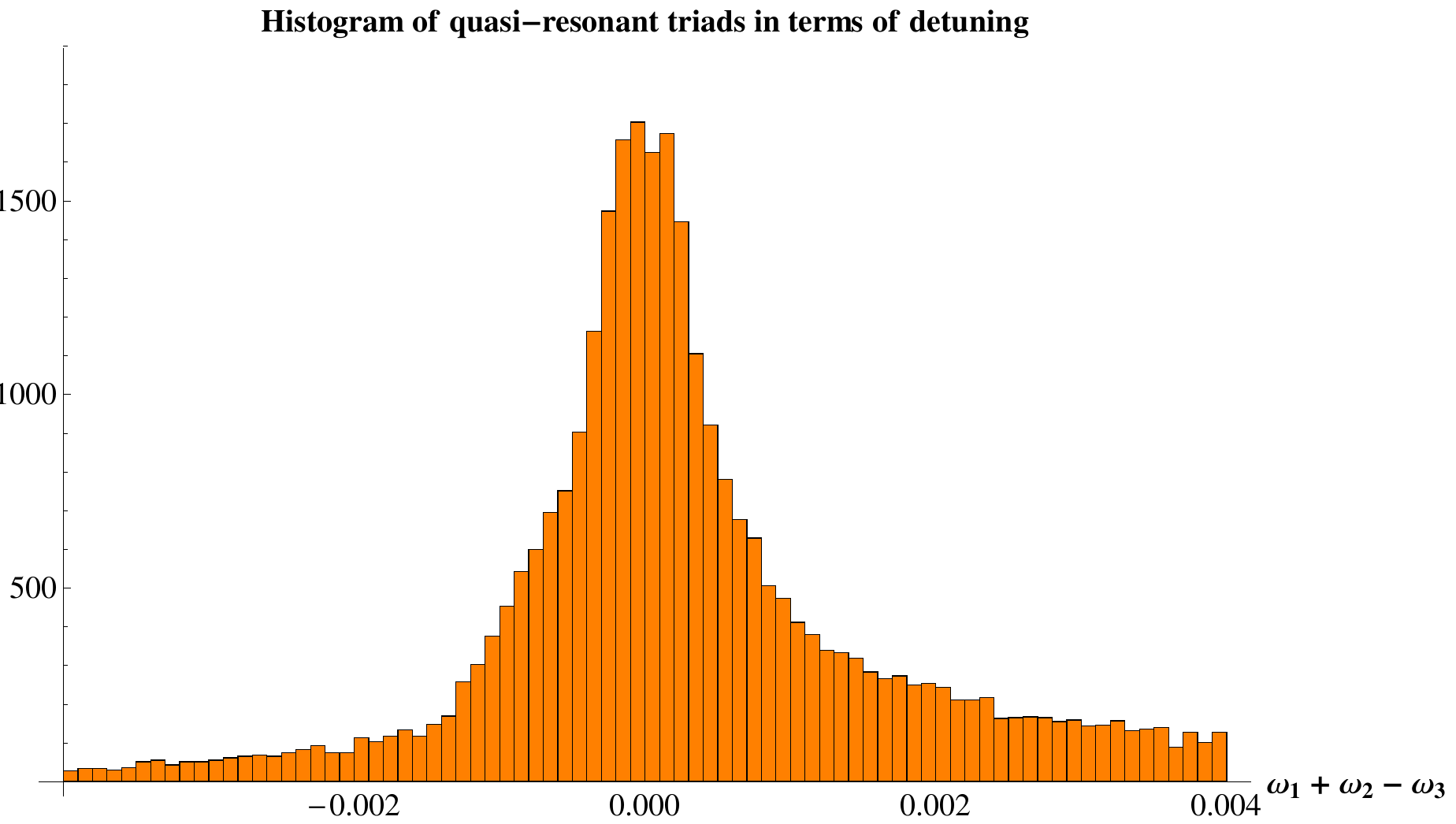}
\includegraphics[width=65mm]{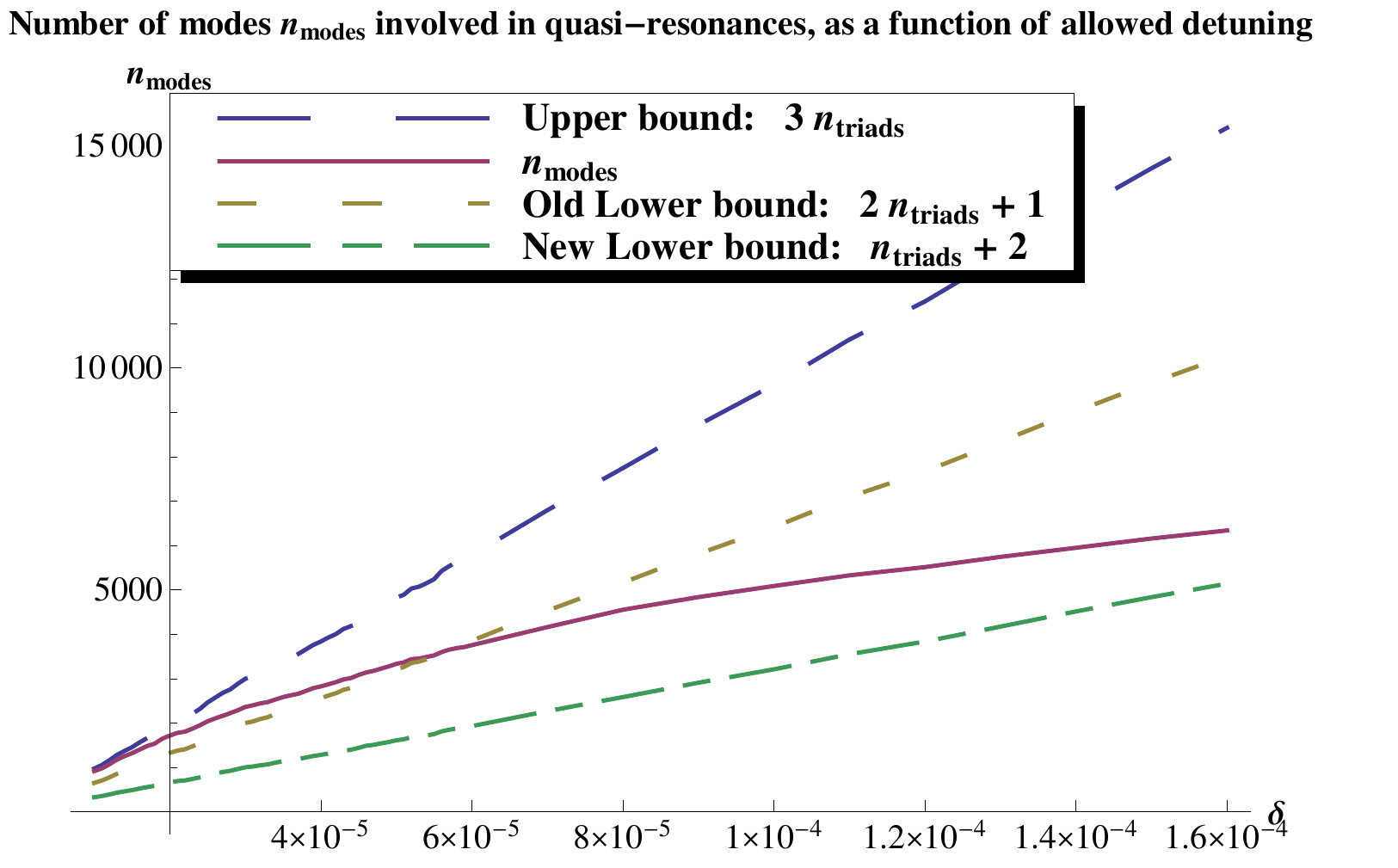}
\caption{\label{fig:qreso_clusters_kite} \textbf{Left Panel:} Histogram, in terms of detuning or frequency mismatch, of the $40434$ quasi-resonant triads generated within the box of size $L=100$ using the analytical method described in Section \ref{sec:quasi}. The histogram appears asymmetrical because of the convention we use to store the representative triads ($0 < k_1 \leq k_2 \leq k_3$). The physically sensible histogram is the symmetrisation of the histogram in the figure, about zero detuning. \textbf{Right Panel:} With respect to the $40434$ quasi-resonant triads within box size $L=100,$ just described in this caption, number of modes $n_{\mathrm{modes}}$ involved in quasi-resonances as a function of the allowed detuning $\delta$ (solid curve, magenta online). For small values of delta, the number of modes is bounded between the upper bound $3 \, n_{\mathrm{triads}}$ (long-dashed curve, blue online) and the lower bound $2 \, n_{\mathrm{triads}} + 1$ (short-dashed curve, yellow online), corresponding to at most one common-mode connectivity type of triads. As $\delta$ grows beyond $5 \times 10^{-5}$, the number of modes goes below the previous lower bound because two-common-mode connectivity type between triads begins to dominate. In this new type of connectivity, the lower bound is $n_{\mathrm{triads}} + 2$ (dot-dashed curve, green online). }
\end{center}
\end{figure}

\noindent \textbf{Numerical Results.} We provide a computation of a sample of the quasi-resonant triads within the box of size $L=100.$ Although for this box size the problem can still be treated by brute-force search (there are only $L^4=10^8$ triads in total), the choice of a small box size allows us to illustrate effectively the main results in terms of connectivity of the quasi-resonant clusters. We start generating the irreducible resonant triads generated by the prime expansions of $N = 4 \times p_j \times q_k^2 \times r_l^2,$ with $j=1, \ldots, 8, \quad k = 1, \ldots, 4 \quad$ and $l=1,\ldots,4,$ combined appropriately with the 2 Pythagorean triples generated by $u = 2, v=\pm 1.$ These are $14854$ irreducible triads in total (computation time = $4$ seconds). Some of these triads are within the box of size $L=100$, but the majority are outside. We pick at random, a small sample of $40$ triads out of the $14854$ irreducible triads, so that the sample is symmetric under mirror symmetry $k_y \to -k_y$ (this is in order to allow for eventual connections of clusters with their mirror images). We apply to these $40$ triads the re-scaling algorithm explained above to generate quasi-resonant triads. As a result, $40434$ quasi-resonant triads within the box are formed. Their distribution in terms of the values of frequency mismatch is plotted in figure \ref{fig:qreso_clusters_kite}, left panel. The histogram appears uneven only because of the convention we use to store the representative triads ($0 < k_1 \leq k_2 \leq k_3$), and physically the distribution must be understood as the symmetrisation of the histogram. The corresponding group of clusters obtained is impossible to plot symbolically in a clear way, because $40434$ connected modes are involved. However, we can study quantitatively the change of the connectivity properties of the clusters as a function of the allowed detuning level. Consider figure \ref{fig:qreso_clusters_kite}, right panel (which is analogous to figure  \ref{fig:reso_clusters}, lower panel, except that now the independent variable is the detuning level $\delta$). For small detuning, below $\delta = 4 \times 10^{-5},$ the connections are still predominantly between isolated triads and connected clusters with \emph{one-common-mode} connectivity. However, the clusters formed when $\delta$ goes beyond about $5\times 10^{-5},$ violate the lower bound that assumes that connectivity, so we deduce that \emph{two-common-mode} connectivity is starting to prevail. It is easy to show that this new connectivity has a lower bound for the number of modes, corresponding to the linear relation $n_{\mathrm{modes}} = n_{\mathrm{triads}} + 2.$ So, the conclusion is that the clusters that form when allowed detuning is high enough, show predominantly the new type of triad connectivity.\\

\begin{figure}
\begin{center}
\includegraphics[height=40mm]{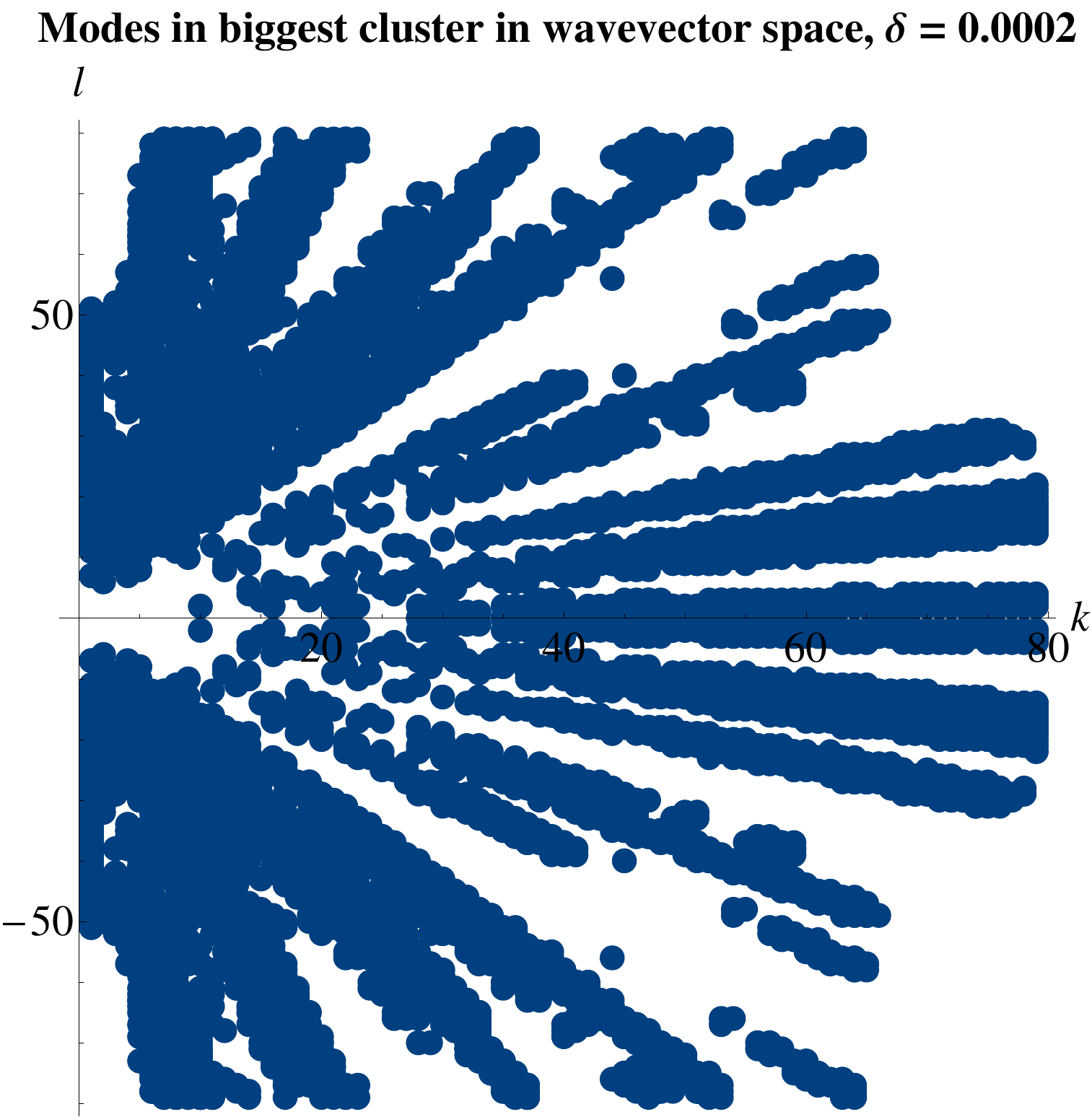}
\includegraphics[height=40mm]{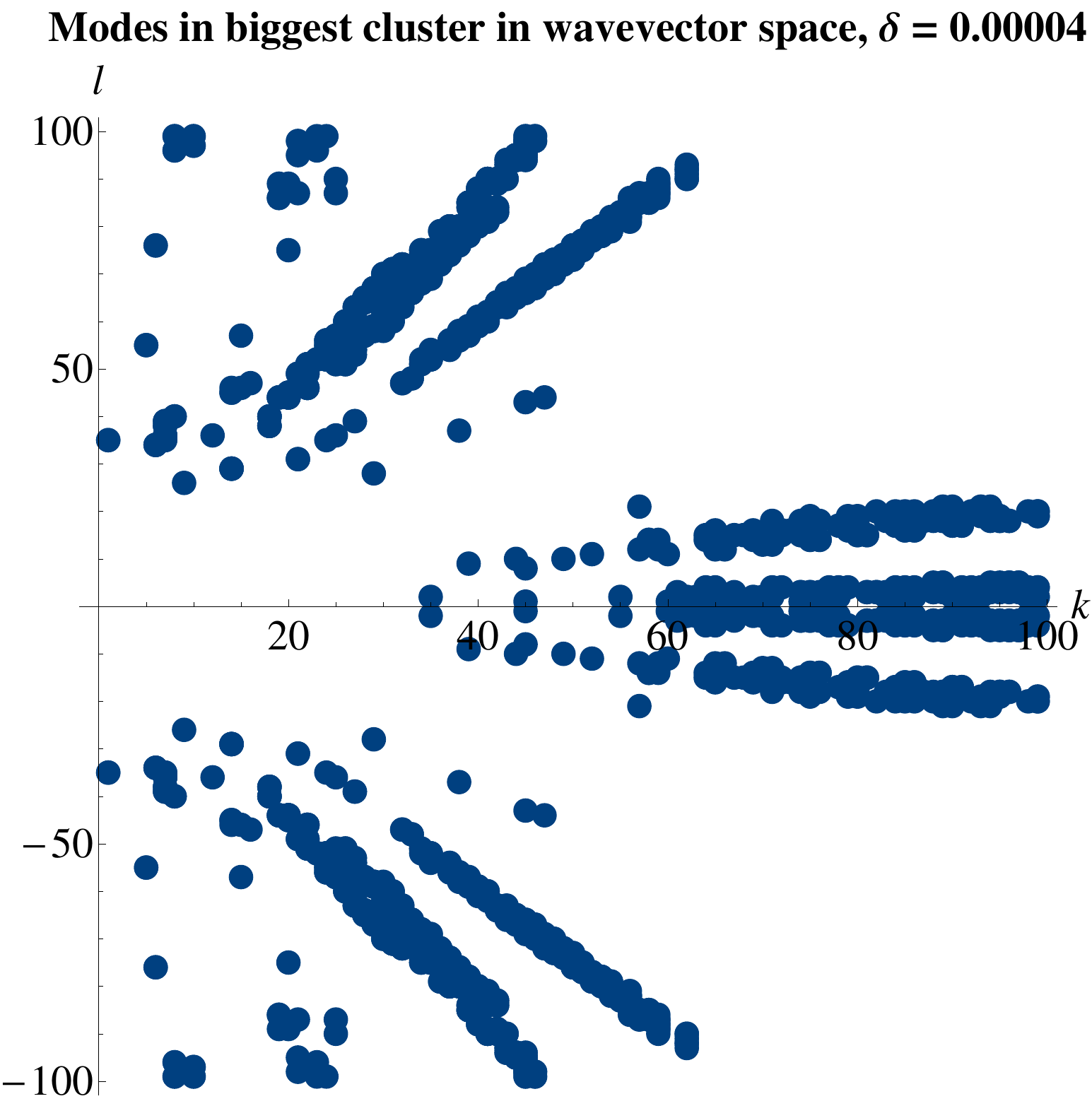}
\includegraphics[height=40mm]{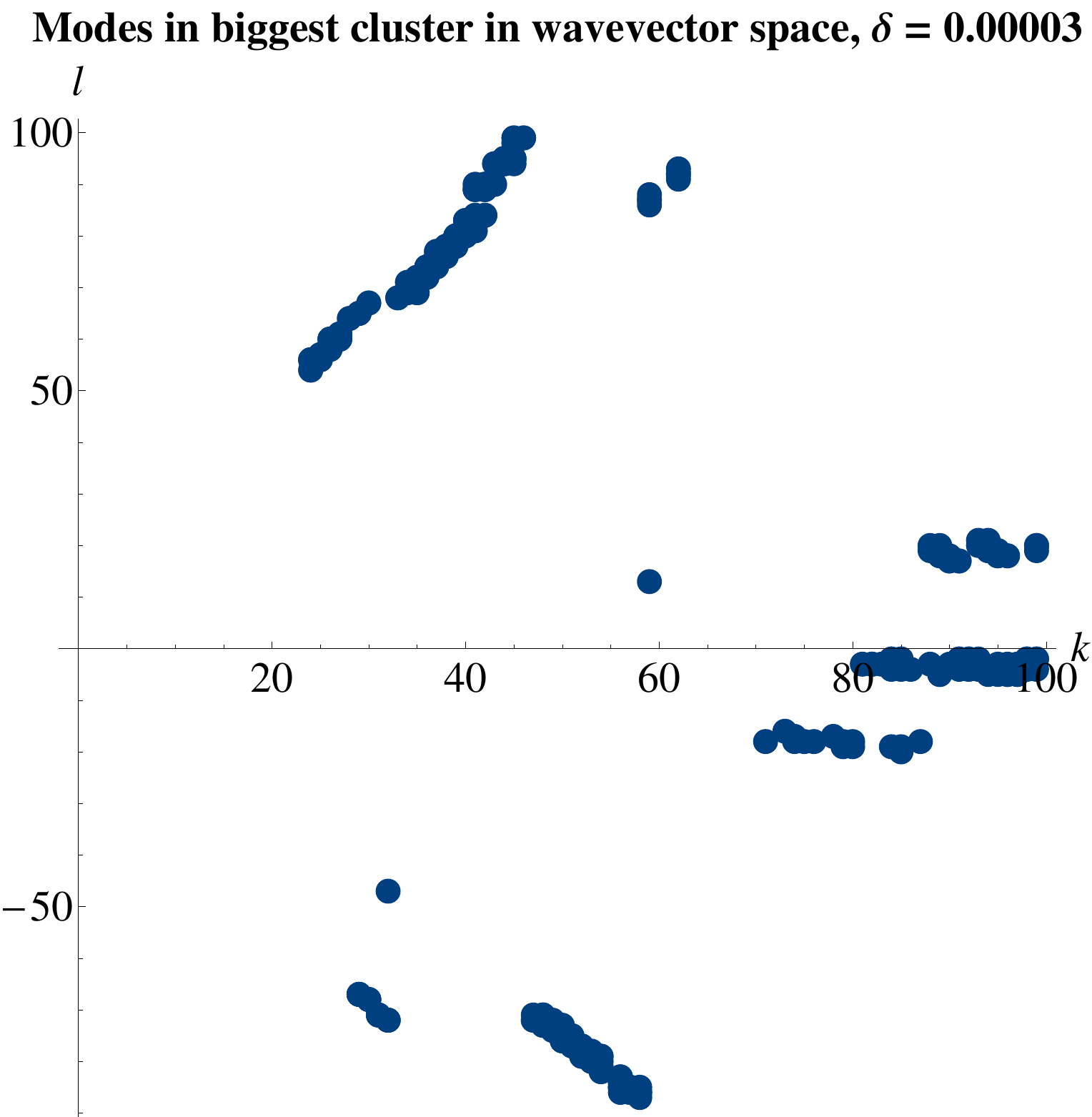}
\caption{\label{fig:big_cluster_qreso} In terms of the $40434$ quasi-resonant triads generated within the box of size $L=100$ using the analytical method described in Section \ref{sec:quasi}: Plots, in $(k,l)$ wavevector space, of the modes in the biggest cluster obtained when the allowed detuning is set to $\delta = 2 \times 10^{-4}$ (left panel), $\delta = 4 \times 10^{-5}$ (middle panel) and $\delta = 3\times 10^{-5}$ (right panel).}
\end{center}
\end{figure}

\noindent \textbf{Onset of turbulence.} A plot, in $(k,l)$ wavevector space, of the modes in the biggest cluster obtained when the allowed detuning is $\delta = 2 \times 10^{-4},$ is shown in figure \ref{fig:big_cluster_qreso}, left panel. For reference, in figure \ref{fig:qreso_clusters_kite}, right panel, such value of $\delta$ would correspond to a high predominance of two-common-mode connections. In figure \ref{fig:big_cluster_qreso}, middle panel, we plot the biggest cluster when detuning is  $\delta = 4 \times 10^{-5}$ and in right panel, we plot the biggest cluster when $\delta = 3\times 10^{-5}.$ It is evident that as the allowed detuning decreases, only high wavenumbers in size are allowed to interact and the angular distribution of active modes becomes more anisotropic.

%In the right panel, for the same value $\delta = 0.0002,$ we plot as a function of box size $L,$ the size of the biggest cluster when we restrict connections to within that box. It is seen that, after a flat region, there is a steep increase (perhaps quadratic) of the number of modes in the biggest cluster as a function of the box size. It is not possible to infer from these plots what are the types of connections between triads.

The previous analyses showed that, as the allowed detuning goes beyond some threshold, the quasi-resonant triads tend to get connected into one big cluster, with connectivity that shifts gradually from a one-common-mode regime to a two-common-mode regime. Let us accept the hypothesis that, in a statistically invariant physical system, the typical amplitude of oscillations $\sqrt{<\psi(x,y,t)^2>}$ (averaged over $x,y,t$) activates quasi-resonant triads with allowed detuning $\delta \propto \sqrt{<\psi(x,y,t)^2>}.$ Evidence in favour of this hypothesis is given by a scaling argument in CHM equation (\ref{eq:CHM}), where the amplitude scales as the inverse of time. Therefore the detuning scales as the amplitude. We deduce that if, in a physical system, we increase gradually the amplitude of the oscillations (say, via forcing or via manipulating the initial conditions), then there is a threshold amplitude below which energy transfers throughout the spectrum of scales and directions are not permitted, and above which they are permitted and in fact favoured by the fact that the connectivity is via two common modes.\\

\begin{figure}
\begin{center}
\includegraphics[width=60mm]{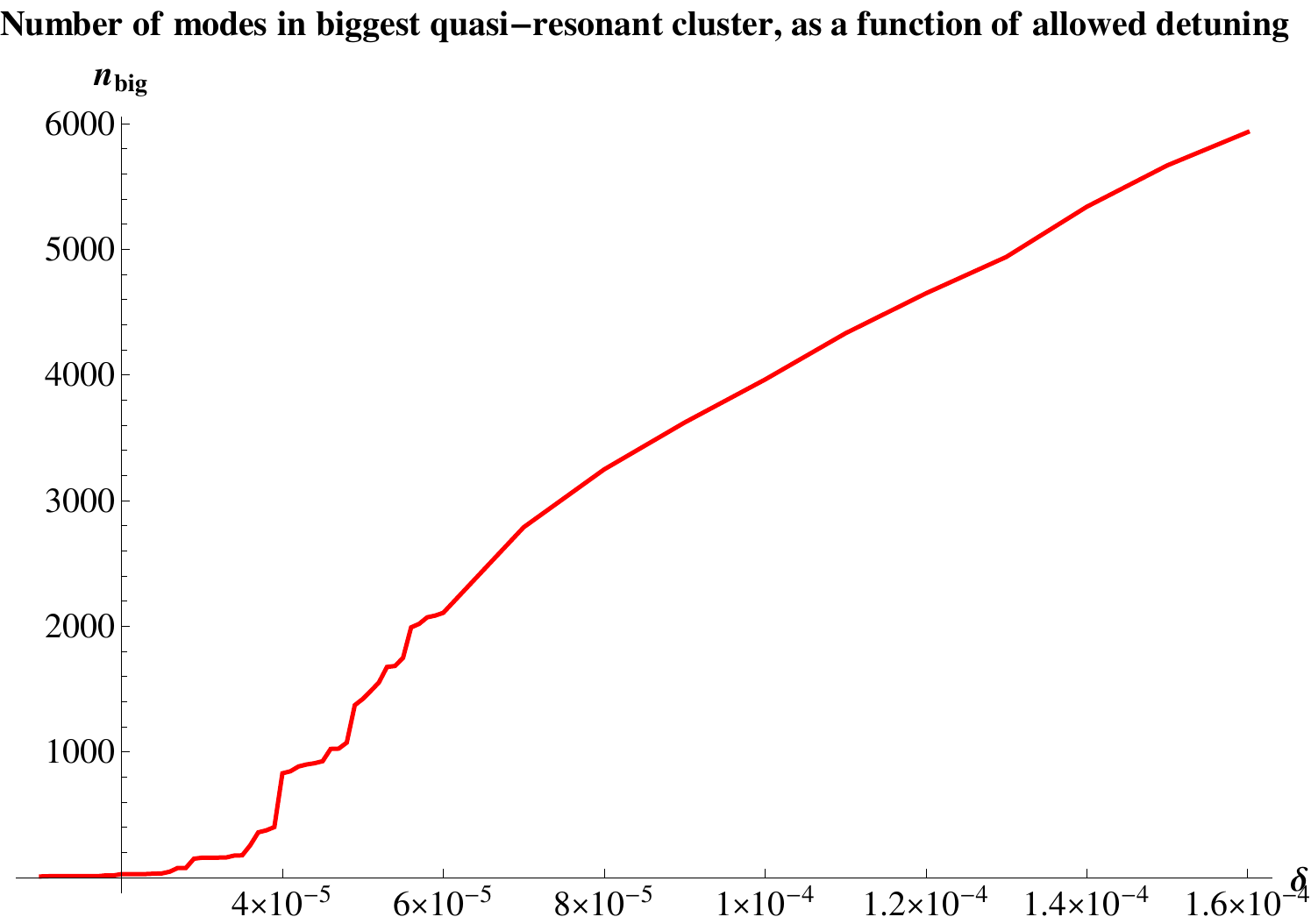}
\includegraphics[width=60mm]{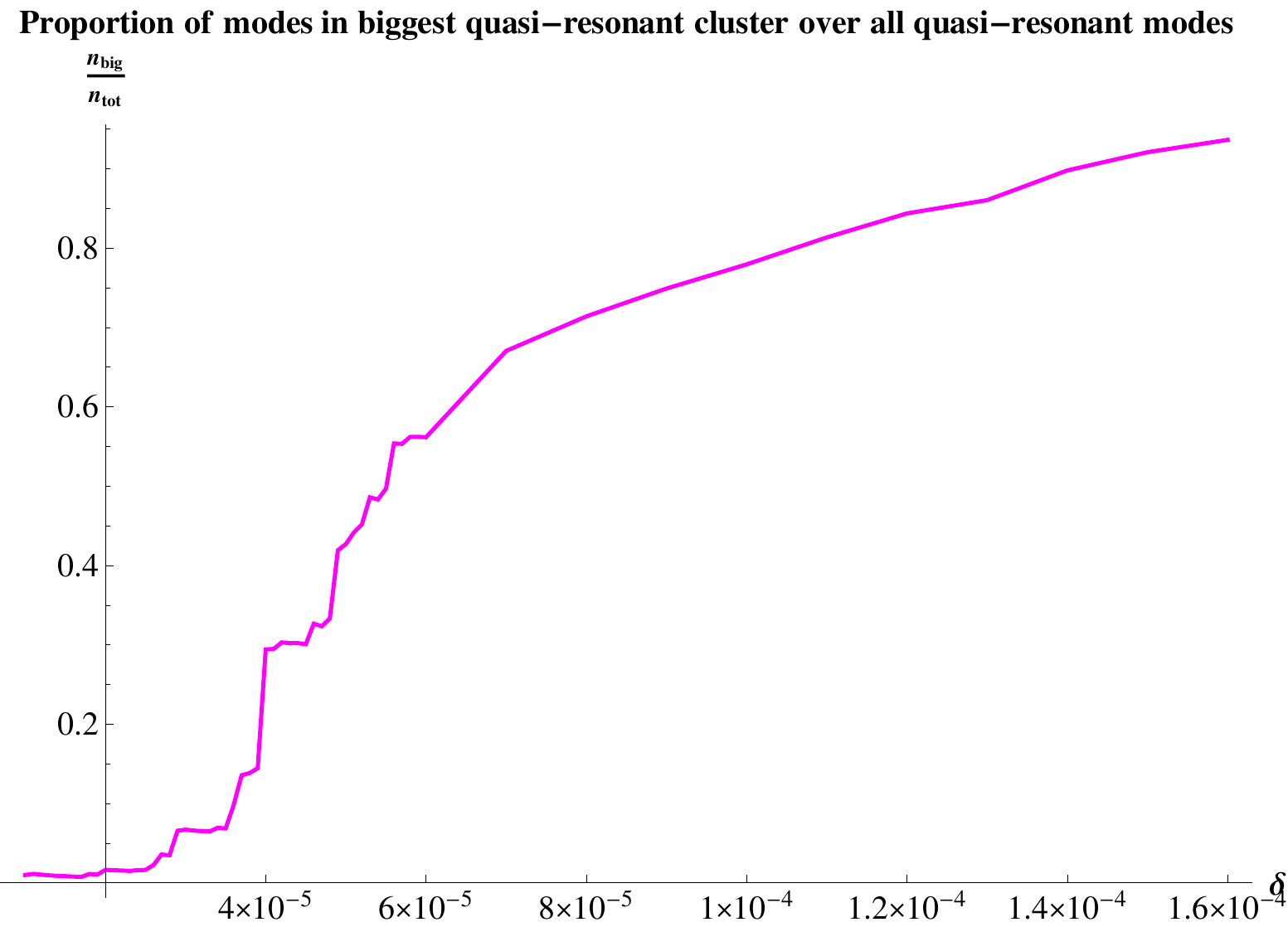}
\includegraphics[width=60mm]{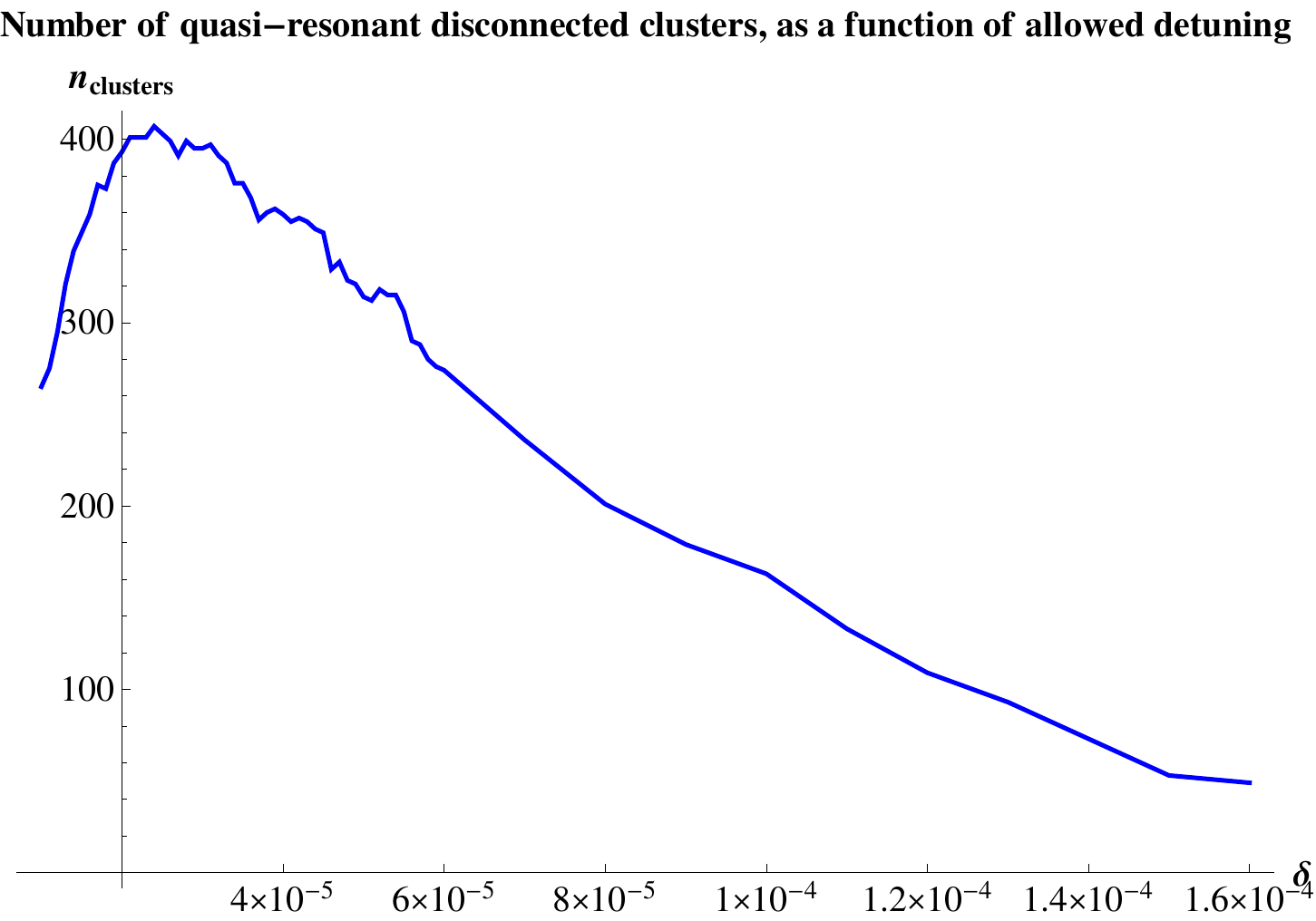}
\caption{\label{fig:qreso_clusters} In terms of the $40434$ quasi-resonant triads generated within the box of size $L=100$ using the analytical method described in Section \ref{sec:quasi}: \textbf{Top Left Panel:} number of modes in the biggest cluster, as a function of the allowed detuning. \textbf{Top Right Panel:} ratio between the modes in the biggest cluster and the total number of modes in all the clusters, as a function of the allowed detuning. \textbf{Lower Panel:} Number of disconnected clusters, as a function of the allowed detuning. The three plots show evidence of a percolation transition at $\delta \approx 2.5\times 10^{-5}$ (see \cite{Ha12a}).}
\end{center}
\end{figure}

\noindent \textbf{Percolation phenomena.} Although we have generated only a subset of the total triads in the box of size $100$ (there are $10^8$ triads in total, and we have $4\times10^4$), the connectivity properties of our clusters behave qualitatively as those of the full set of triads. In a paper in preparation (in collaboration) by one of the authors \cite{Ha12a}, a rigorous computation of all triads in a box of size $256$ gives indication of percolation phenomena. Here we show that similar behaviour is observed in our clusters. In figure \ref{fig:qreso_clusters}, top left panel, the size of the biggest cluster as a function of the detuning level $\delta$ is plotted. We see that there is an interesting transition to predominance of the big cluster, at a certain value of detuning $\delta.$ More evidence of this transition is given in figure \ref{fig:qreso_clusters}, top right panel, where the ratio between the size of the biggest cluster and the total number of modes involved is plotted. It can be inferred from the plot, that at large enough values of $\delta,$ the big cluster contains $95\%$ of the modes involved. Another piece of evidence is given in figure \ref{fig:qreso_clusters}, lower panel, where the total number of disconnected clusters is plotted as a function of detuning level. The maximum in this figure explains the transition in terms of connectivity: when $\delta$ goes beyond $2.5 \times 10^{-5},$ the disconnected clusters begin to connect at a rate that is faster than the rate of appearance of new disconnected clusters.

\section{New Developments: Aspect ratio not equal to one}
\label{sec:new_dev}
In this section we consider briefly the case when aspect ratio is not equal to $1$. In other words we will be considering the case where

$$\omega(k,l) = -\frac{\beta\,k}{k^2 + f^2l^2}\,,$$
where $f=f_{1} \sqrt{ f_{2} }$, $f_{1} \in \mathbb{Q}$ and $f_{2} \in \mathbb {N}$ square free. The choice of $f$ so that $f^2$ is rational is important: otherwise the resonance condition analogous to equation \eref{eq:simplif} would split into two or more independent equations, depending on the algebraic degree of $f$. For example, if $f$ was transcendental then all non-zero solutions of the resonance conditions would involve zonal modes (e.g., $k_3=0$) which do not interact physically in an exact resonant triad.

The analysis and transformations for the case $f^2 \in \mathbb{Q}$ are analogous to the case $f=1$, with some minor changes. We omit the intermediate steps and just present the final form of the mappings. With the assumption $k_3, k_1 \neq 0,$ we define
$$D   = \frac{k_3}{k_1} \, \frac{k_3 k_1 + f^2l_3 l_1}{k_3^2+f^2l_3^2}$$
and we have two cases:

\subsection{Case $D=0$}
In this case, the ratio $\frac{l_3}{f^2k_3}$ must be a pure-cube rational and we obtain a triad of the form $(k_1,l_1), (k_2,l_2), (k_3,l_3)$ with
\begin{equation}
\label{eq:cube_solnf}
 (k_1,l_1) = -\left(\frac{l_3}{f^2 k_3}\right)^{1/3}\,(-f^2 l_3,k_3), \quad (k_2,l_2) = (k_3 - k_1,l_3-l_1),
\end{equation}
where an overall scaling factor might be needed in order that all wavenumbers be integer.

\subsection{Case $D\neq 0$}

Provided the ratio $l_3/f^2k_3$ is not a pure-cube rational, we can produce a rational bijective mapping from an integer triad $(k_1,l_1), (k_2,l_2), (k_3,l_3)$ satisfying all resonance conditions, to the variables of an elliptic curve
\begin{equation}
\label{eq:ellicurf}
 X^3 - 2\,D\,X^2 + 2\,D\,X - D^2 = f^2 Y^2\,.
\end{equation}
The mapping, bijective up to re-scaling of the triad wavenumbers, is given explicitly by:
\begin{equation}
 \label{eq:mapf}
X = \frac{k_3}{k_1} \, \frac{k_1^2+f^2l_1^2}{k_3^2+f^2l_3^2}\,, \quad Y = \frac{k_3}{k_1} \, \frac{k_3 l_1 - k_1 l_3}{k_3^2+f^2 l_3^2}\,, \quad D = \frac{k_3}{k_1} \, \frac{k_3 k_1 + f^2l_3 l_1}{k_3^2+f^2l_3^2},
\end{equation}
and with inverse
\begin{equation}
 \label{eq:map_invf}
\frac{k_1}{k_3} = \frac{X}{D^2+f^2 Y^2}\,,\quad \frac{l_1}{k_3} = \frac{X}{f^2 Y} \left(1-\frac{D}{D^2+f^2 Y^2}\right)\,,\quad \frac{l_3}{k_3} = \frac{D-1}{f^2 Y}\,,
\end{equation}
and $(k_2,l_2) = (k_3-k_1,l_3-l_1).$

\subsection{Classification of solutions of triad equations in terms of Fermat's theorem of sums of squares (case $D\neq0$)}

We consider the case $D \neq 0$ described in the previous Subsection.

  We can rewrite the elliptic curve \eref{eq:ellicurf} as:
\begin{equation}
 \label{eq:ellicur1f}
f^2\, {Y}^2 + \left(D + X^2 - X\right)^2 = X^2 \left(X^2-X+1\right)\,,
\end{equation}
and we can divide by $X$ because $X \neq 0\,,$ obtaining:
\begin{equation}
 \label{eq:ellicur2f}
f^2 \,\left(\frac{{Y}}{X}\right)^2 + \left(\frac{D}{X} + X - 1\right)^2 = X^2-X+1\,.
\end{equation}
The last expression is a quadratic form that is best written in diagonal form by defining
\begin{equation}
\label{eq:Xmnf}
X \equiv -\frac{m+n}{m-n}, \quad m, n \in \mathbb{Z}\,,
\end{equation}
so we get the equation:
\begin{equation}
 \label{eq:ellicur3f}
f^2 \, \left(\frac{{Y} (m-n)^2}{m+n}\right)^2 + \left(\frac{D(m-n)^2}{m+n} + 2\,m\right)^2 = 3\,m^2 + n^2\,.
\end{equation}
Finally, using the definition of the aspect ratio $f = f_1 \sqrt{f_2}$ with $f_1 \in \mathbb{Q}$ and $f_2 \in \mathbb{N}$ square free, we obtain the equation:
\begin{equation}
 \label{eq:ellicur4f}
f_2 \, \left(\frac{{f_1 \, Y} (m-n)^2}{m+n}\right)^2 + \left(\frac{D(m-n)^2}{m+n} + 2\,m\right)^2 = 3\,m^2 + n^2\,.
\end{equation}
The problem is thus reduced to finding representations of integers as sums of the form $3 m^2 + n^2$ with $m,n$ integers, and of the form $f_2 s^2 + q^2$ with $s,q$ rationals, where $f_2$ is a square-free natural number. The solution to this problem depends of course on the explicit value of $f_2,$ but the method is straightforward and computable, due to the so-called Hasse-Minkowski theorem, see \cite [pages 61-69]{Shaf}. Essentially, the solution algorithm will be very similar to the one shown in Section \ref{sec:Fermat}. The details will be presented in a forthcoming paper.

\section{Applicability of the method to the general CHM equation}
\label{sec:concl}
In this Section we consider briefly some remarks about the search for exact resonances in the case of the Charney-Hasegawa-Mima equation with arbitrary coefficient $F>0.$ Our paper focused on the case $F=0$ due to a technical reason: in the case $F>0,$ the system of Diophantine equations stemming from the resonance conditions becomes more difficult to analyse in terms of our mappings. We believe that the general case $F>0$ is far from obvious and requires more research for its resolution. To provide evidence of this, we show three results obtained by direct computations on a box of wavenumbers $0 < k \leq 60, \,\, -60 \leq l \leq 60.$

\begin{figure}
\begin{center}
\includegraphics[width=60mm]{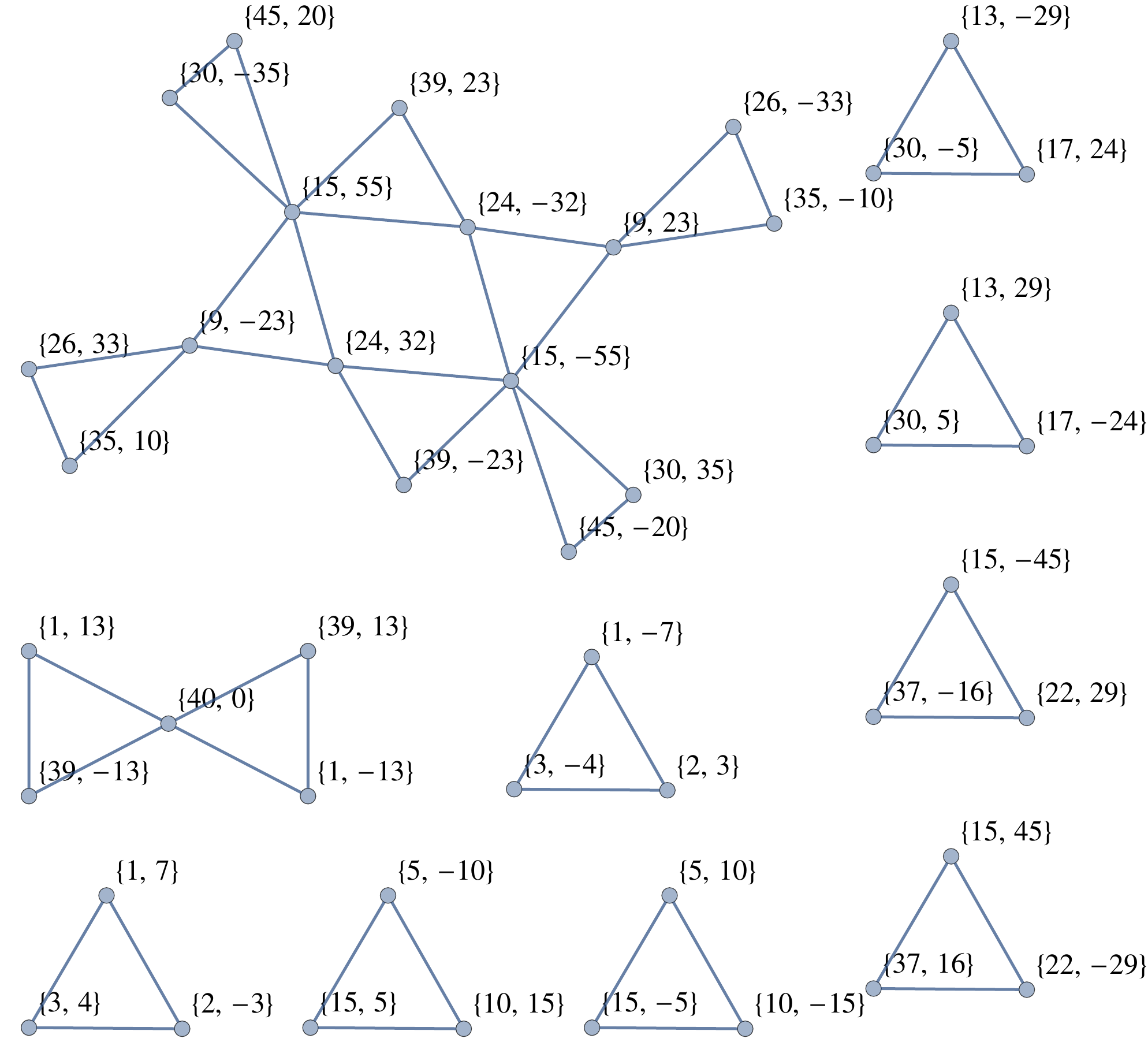}
\caption{\label{fig:F_875_clusters_L_60} Exact resonant clusters for the CHM equation, dispersion relation (\ref{eq:dispersionF}) with $F=875,$ in the domain $|k|, |l| \leq 60.$ }
\end{center}
\end{figure}

First, the resonance conditions for $F>0$ are again Eqs. (\ref{eq:reso_1})--(\ref{eq:reso_3}), but the frequency is given by the following dispersion relation:
    \begin{equation}
\label{eq:dispersionF}
     \omega(k,l)  \equiv -\frac{\beta \, k}{k^2+l^2+F}\,.
     \end{equation}
 After some simple algebra, it follows that the resonance conditions imply that $F$ must be rational, and in fact it must be given by:
 \begin{equation}
 \label{eq:Fsol}
  \scriptstyle F = -\frac{{k_2} \left({k_1} ({k_1}+{k_2}) \left({k_1}^2+{k_1} {k_2}+{k_2}^2\right)+2 {k_1} ({k_1}+{k_2}) {l_1}^2+{l_1}^4\right)+2 {k_2} {l_1} \left({k_1} ({k_1}+{k_2})+{l_1}^2\right) {l_2}+2 {k_1} {k_2} ({k_1}+{k_2}) {l_2}^2+2 {k_1} {l_1} {l_2}^3+{k_1} {l_2}^4}{{k_2} \left(3 {k_1} ({k_1}+{k_2})+{l_1}^2\right)+2 ({k_1}+{k_2}) {l_1} {l_2}+{k_1} {l_2}^2}.
 \end{equation}
Keeping in mind that $k_1,l_1,k_2,l_2$ are integers, a brute-force search is possible where we cover all possible triads in wavenumber space and only keep the triads that produce $F>0$ according to equation (\ref{eq:Fsol}). On the given box of size $60,$ there are $10614734$ triads with generic interaction coefficients (i.e., such that no interaction coefficient is identically zero), and amongst these triads, only $633360$ triads (about $6\%$ of the total) correspond to exact triads for some positive, and necessarily rational, $F$. Of these exact triads, only $22460$ triads correspond to \emph{integer} values of $F.$ Despite the relatively small fractions, the number of integer cases is big enough to give us some hope of finding, in the future, a generalisation of the method valid for $F=0.$ However, there is an important observation: let us order the values of $F$ in terms of the total number of resonant triads that they give rise to, within the box considered. These values of $F$ do not show an obvious pattern, other than the fact that the majority of them are integers. The first few values are: $F=0$ ($28$ triads), $F=175$ ($26$ triads), $F=10$ ($22$ triads), the cases $F=250, 1075, 50, 475$ all give $20$ triads each, and so on. But not all interesting $F$ values are multiples of $5$ or integers. For example, $F=986$ gives $16$ triads and $F=590/3$ gives $12$ triads. In terms of connectivity, most of the cases with more than $10$ triads give groups of clusters (with one-common-mode connectivity), of particular interest being the case $F=875,$ which has $18$ triads, subdivided into $8$ isolated triads, $1$ butterfly and $1$ cluster of $8$ triads (see figure \ref{fig:F_875_clusters_L_60}).

\begin{figure}
\begin{center}
\includegraphics[width=60mm]{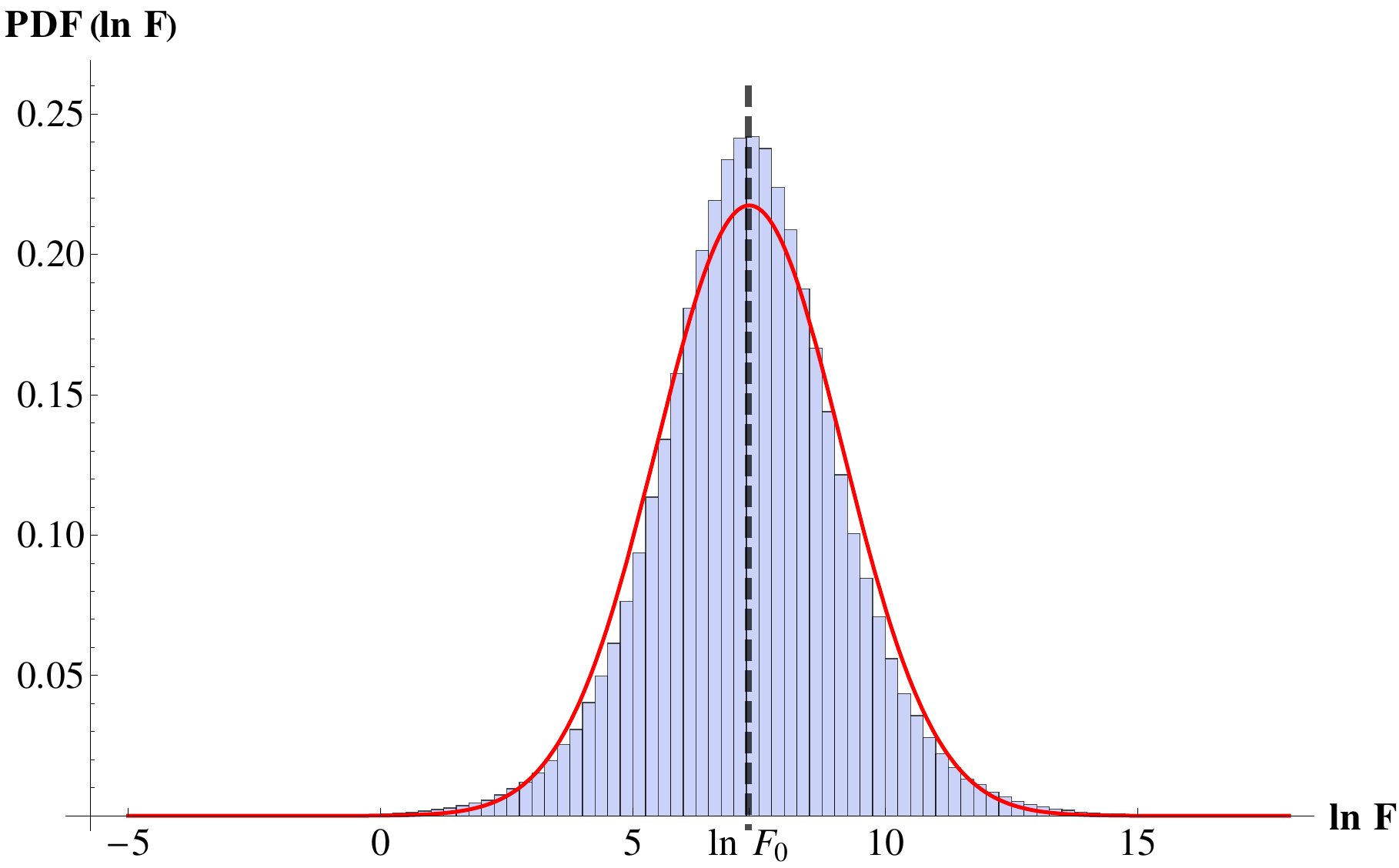}
\includegraphics[width=65mm]{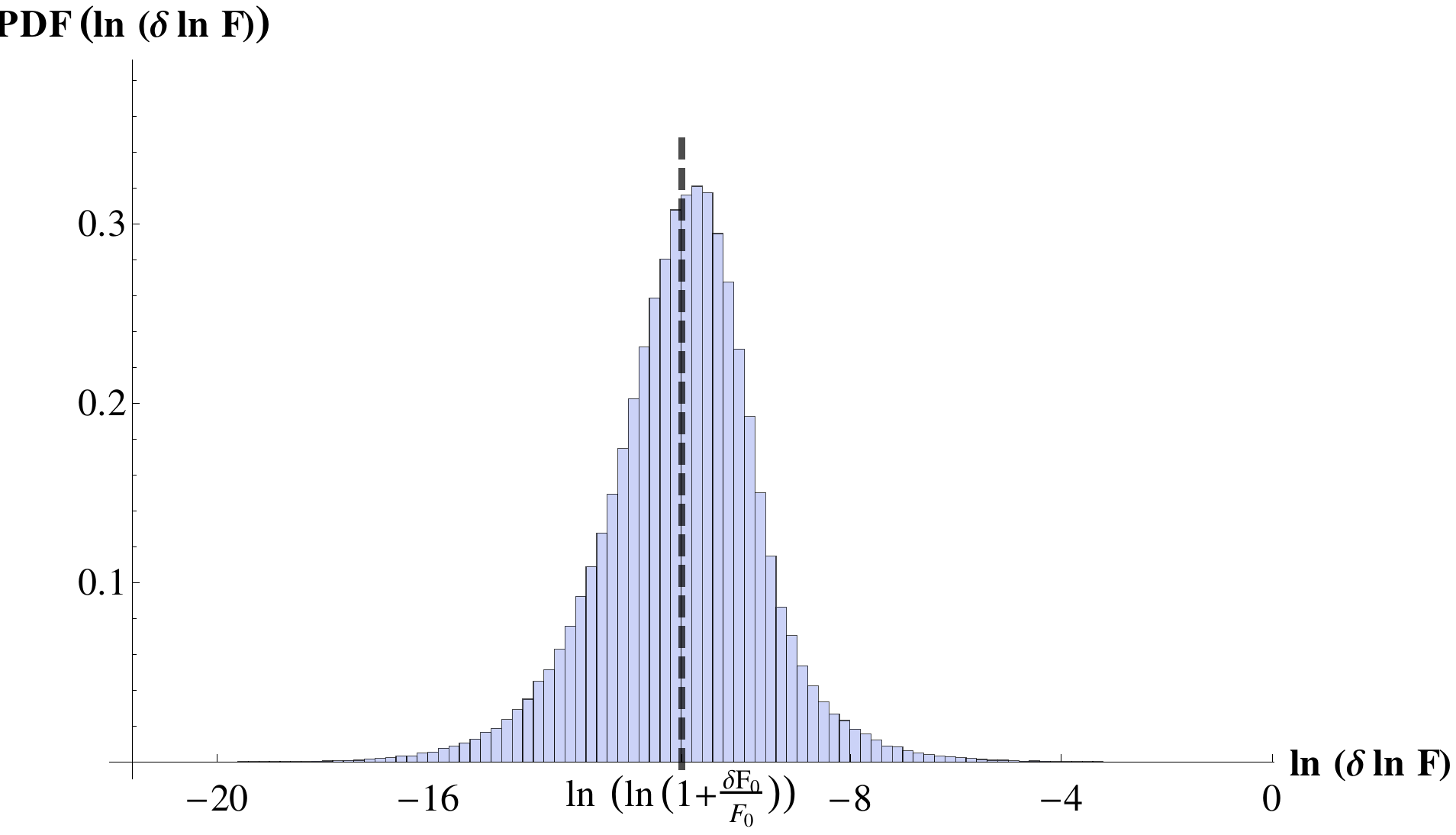}
\caption{\label{fig:lnF_distr_L_60} \textbf{Left Panel:} Bars: probability density function (PDF) of the natural logarithms of the $308544$ positive values of $F$ obtained by evaluating Eq.(\ref{eq:Fsol}) over the grid $1 \leq k_1, k_2 \leq 60,$ $0 \leq |l_1|, |l_2| \leq 60.$ Solid curve: Gaussian distribution with the same mean and variance as the PDF. Vertical dashed line: mean of the distribution, $\ln F_0 \approx 7.31.$ \textbf{Right Panel:} Bars: PDF of the natural logarithms of the separation between contiguous values of $\ln F,$ taken from the $308544$ positive values of $F$ used in left panel. Vertical dashed line: mean of this distribution, $\ln (\ln (F_0+\delta F_0)-\ln F_0) \approx -11.2.$}
\end{center}
\end{figure}

As a second insight, we look at the distribution of the $308544$ different values of $F > 0$ found numerically, that give rise to exact resonant triads within the box of size $L = 60.$ It turns out that these values are distributed in a kind of Log-Normal distribution (perhaps with fat tails). This is evidenced by looking at figure \ref{fig:lnF_distr_L_60}, left panel, which shows the probability density function (PDF) of the natural logarithms of the $308544$ values of $F$ (bars). The shape is quite symmetric. For comparison, a Gaussian distribution with the same mean and variance is shown (solid curve), and a vertical dashed line denotes the centre of the distribution, $\ln F_0,$ the typical value of $\ln F.$ There is an obvious mismatch between the histogram and the Gaussian fit, particularly near the centre of the distribution. We have observed (figure not shown) that the position of the centre of the distribution seems to follow the pattern $F_0 \approx 0.5 \,L^2,$ where $L$ is the box size ($L=60$ for figure \ref{fig:lnF_distr_L_60}, giving $F_0 \approx 1489$). Physically this means that most of the resonant triads found have values of $F$ comparable to the typical squares of the wavevectors, so that the terms in the denominator of equation (\ref{eq:dispersionF}) become balanced.

As a third insight, the fact that the values of $F$ are rational imply that they are somehow scattered. A measure of this scattering is obtained by looking at the distribution of $\delta \ln F,$ the separation between contiguous values of $\ln F$ in our list of $308544$ values. The logarithm of the separation is distributed according to the PDF shown in figure \ref{fig:lnF_distr_L_60}, right panel. The distribution is asymmetric. The mean of this distribution (in other words, the typical logarithm of the separation) seems to decrease with increasing box size (figure not shown). For $L=60,$ we conclude that contiguous values of $F$ near $F_0$ are separated by a typical distance of $\delta F_0 \approx 0.0205.$ \\

\noindent \textbf{Quasi-Resonances for $F > 0$.} It is possible to show that the $F=0$ resonant triads obtained with our method can be used directly as quasi-resonant triads in the case $F \gtrsim 0,$ via the replacement $k^2 + l^2 \to k^2 + l^2 + F$ in the denominator of the dispersion relation (\ref{eq:dispersion}). When $\sqrt{F}$ is comparable to or greater than the box size under consideration, our $F=0$ method will have to be modified, because it relies on a down-scaling of the wavenumbers in the triads, a procedure that is not possible in the case $F>0$ without re-scaling $F$ itself: a direct application of our method in the general case $F>0$ will miss some triads, and will produce too high detuning levels.

Let us assume we have some general method for constructing the exact resonances for any rational $F > 0.$ Then one can devise a method for constructing quasi-resonant triads, with controlled value of detuning, for any real value of $F>0.$ The method is as follows: (i) For any \emph{real} $F_*>0$ there is a neighbourhood of rational values of $F$ near $F_*,$ whose exact triads are in fact quasi-resonant triads for $F_*.$ (ii) Exact resonant triads have values of $F$ distributed in a smooth PDF as in figure \ref{fig:lnF_distr_L_60}, which guarantees that there will be a sufficient number of quasi-resonant triads.

Notice that when $F_*$ is reasonably small (but finite), this new method can be used directly to find quasi-resonant triads via a direct search (using for example the data leading to figure \ref{fig:lnF_distr_L_60}). But when $F_*$ is too large, or when the box size is too large, the direct search method will take too long and one will need to use an analytic method to find the exact resonances for any rational $F$. This research will be part of a subsequent work.

Another interesting question that we will consider in the future is how to extend our method to treat other nonlinear PDE models of turbulence, with dispersion relations that are either polynomials or quotients of polynomials. In particular, we will study the barotropic vorticity equation on the sphere, leading to the so-called Rossby-Haurwitz triads, more suitable for atmospheric models. This equation can be treated using a direct extension of our method.

\section*{Acknowledgements}
We acknowledge useful scientific discussions with C. Connaughton, J. Harris, D. Holmes, P. Lynch, G. McGuire, S. Nazarenko, M. Reid, S. Siksek, B. Quinn.
MDB acknowledges UCD support under project SF564.

\appendix

\section{Numbers of the form $3 \,m^2 + n^2$}

The detailed discussion on this topic is available in \cite[Chapter 1]{Cox} and \cite[Section 7]{Shaf}. We briefly describe Fermat's approach to finding, for a given integer $N,$ all integers $m,n$ such that $N = 3\, m^2 + n^2.$ The basic observation is that all we need is to consider such problem when $N$ is a prime number. Then a general $N$ is written as a product of primes and the corresponding question can be solved by using the Brahmagupta identity iteratively:
\begin{equation}
 \label{eq:Brahma}
(M\, a^2 + b^2)(M \, c^2 + d^2) = M\, (a d + b c)^2  +  (b d - M\,a c)^2 = M\, (a d - b c)^2  +  (b d + M\,a c)^2 \,,
\end{equation}
valid for any square-free $M \in \mathbb{Z}$ and $a,b,c,d \in \mathbb{Q}.$

%It follows from modular arithmetic that a prime number $P$ of the form $P = 3\, m^2 + n^2,$ with $m,n$ integers, must be either equal to $3$ or of the form $3 k + 1,$  where $k$ is some integer. The corresponding values of $m$ and $n$ are unique up to changes of sign. This classifies then all possible primes appearing \emph{with odd powers} in the prime expansion of $N.$  Primes of the form $3 k - 1$ can appear in the expansion, \emph{provided they do so with even powers}. Such even powers have again a unique representation in the ``trivial'' form $n^2$, i.e., one must take $m=0$. Finally, there is a special case which is the representation of the number $4$ as $4 = 3 \times 1^2 + 1^2,$ so $m$ and $n$ can take the values $\pm 1.$ Using the Brahmagupta identity \eref{eq:Brahma} one can see that other powers of $4$ have a ``trivial'' representation ($m=0$).

Fermat's argument shows that a prime $P$ can be written in the form $P = 3\, m^2 + n^2,$ if and only if $P=3$ or $P= 3 k + 1,$  where $k$ is some integer. In such case the coprime integers $m, n$ are unique up to signs. Similarly, any positive integer power of such primes has a unique (up to sign) representation in the form $3 \, m^2 + n^2.$ For example, $7 = 3 \times 1^2 + 2^2,$ $7^2 = 3 \times 4^2 + 1^2$ and $7^3 = 3 \times 9^2 + 10^2$ in unique ways. All these can be obtained from application of the Brahmagupta identity iteratively starting from the expansion for $7 = 3\times 1^2 + 2^2.$ Non-coprime integers are obtained trivially from the coprime solutions. For example, we can also write $7^3 = 7^2 \times 7 = 7^2 \times (3\times 1^2 + 2^2) = 3\times 7^2 + 14^2.$

By allowing for the special case $4 = 3 \times 1^2 + 1^2,$ one can state the result below due to Fermat \cite[Chapter 1]{Cox}:\\

\noindent \textbf{Theorem 1.} A general positive integer $N$ is expressible in the form $N = 3\, m^2 + n^2,$ if and only if the prime expansion of $N$ contains no odd powers of primes of the form $P = 3\,k - 1\,,$ with $k$ integer. It can contain odd powers of $4$, $3$ and primes $P = 3\,k+1.$ Moreover, the ways in which the integer $N$ can be represented as $N = 3\,m^2 + n^2$ with $m,n$ integers, can be computed by repeated iteration of the Brahmagupta identity \eref{eq:Brahma} to each product of primes.\\

For example, the number $N = 4 \times 3 \times 7 = 84$ can be written as $84 = 3 \times 5^2 + 3^2 = 3\times 1^2 + 9^2.$ But neither $N=5$ nor $N = 11$ can be written in such form.\\

\noindent \textbf{Remark.} For our computational purposes, for a given positive integer $N$ according to Theorem 1, it is important to have an algorithm to compute all possible values of the integers $m', n'$ in $N = 3\,(m')^2 + (n')^2.$ Such algorithm is based on the complex representation of prime factors $P = 3 \,m^2 + n^2 = (n + \sqrt{3} \,i\,m)(n - \sqrt{3} \,i\,m)\,,$ where $i=\sqrt{-1}$, and is equivalent to the iterative application of Brahmagupta identity \eref{eq:Brahma}.

\section{Numbers of the form $s^2 + q^2$}

This is the best known problem of this type, leading to the so-called Pythagorean triples, see \cite[chapter V]{Daven} and \cite[Chapter VI]{Dickson}. The relevant questions are: \\

\noindent Which positive integers $N$ can be represented in the form $N = s^2 + q^2$, with $s,q$ rationals? In how many ways can we write such representation?\\

In this case $s,q$ can be rationals because there are no restrictions on $Y$ and $D$ in equation \eref{eq:ellicur3}. However, the question in terms of rationals can be reduced to a question in terms of integers, by writing the number $1$ as a sum of squares of two rationals:
\begin{equation}
\nonumber
1 = \left(\frac{2 \,u\,v}{u^2+v^2}\right)^2 + \left(\frac{u^2 - v^2}{u^2+v^2}\right)^2\,, \quad \forall \,u, v \in \mathbb{Z}.
\end{equation}
This equation is easy to verify (directly or via Brahmagupta identity) and it leads to the Pythagorean triples $(2\,u\,v, u^2-v^2,u^2+v^2),$ which are the integer sides of a right triangle. Using equation \eref{eq:elegant1} it is possible to write down a partial solution to the question above, in the form:\\

\noindent \textbf{Theorem 2.} Let $N = s^2 + q^2$ with $N \in \mathbb{Z}$ integer and $s,q \in \mathbb{Q}.$ Then, there exist $u,v \in \mathbb{Z}$ and $S,Q \in \mathbb{Z}$ such that
\begin{equation}
 \label{eq:sq_rep}
 s = \frac{2\,u\,v \,Q + (u^2-v^2)\,S}{u^2+v^2}\,,\quad  q = \frac{- 2\,u\,v \,S + (u^2-v^2)\,Q}{u^2+v^2}\,,
\end{equation}
and $N = S^2 + Q^2.$ \\

Now that the problem has been reduced to finding the integer representations of $N$ in the form $N = S^2 + Q^2,$ we can use Fermat's results:\\

\noindent \textbf{Theorem 3.}\cite[Section I, Chapter V]{Daven} A general positive integer $N$ is expressible in the form $N = S^2 + Q^2,$ if and only if the prime expansion of $N$ contains no odd powers of primes of the form $P = 4\,k - 1\,,$ with $k$ integer. It can contain odd powers of $2$ and primes $P = 4\,k+1.$ Moreover, the ways in which the integer $N$ can be represented as $N = S^2 + Q^2$ with $S,Q$ integers, can be computed by repeated iteration of the Brahmagupta identity \eref{eq:Brahma} to each product of primes.\\

\noindent \textbf{Numbers of the form $3 \,m^2 + n^2$ and $s^2+q^2.$} Combining the results encapsulated in Theorems 1, 2 and 3, we obtain the following\\

\noindent \textbf{Corollary 4.} A general positive integer $N$ is expressible in the forms $N = 3 \,m^2 + n^2 = s^2+q^2$ with $m,n \in \mathbb{Z}$ and $s,q \in \mathbb{Q},$  if and only if the only odd powers of primes appearing in the prime expansion of $N$ are those of the primes of the form $P = 12 \, k + 1\,,$ where $k$ is some integer. As a special case, odd powers of $4$ are also allowed in the prime expansion of $N$. Moreover, the different solutions for $m,n$ and $s,q$ can be fully classified in terms of the basic representations of the prime factors in $N$ and the Pythagorean triples appearing in equation \eref{eq:elegant1}.\\

For example, the integer $N=13$ is expressible as $N = 3\times 2^2 + 1^2 = \left(\frac{2\,u\,v \,Q + (u^2-v^2)\,S}{u^2+v^2}\right)^2 + \left(\frac{- 2\,u\,v \,S + (u^2-v^2)\,Q}{u^2+v^2}\right)^2\,,$ with $(S,Q) = (3,2)$ and $u,v$ arbitrary integers.\\

%\section*{References}
\bibliographystyle{elsarticle-num}

\bibliography{Paper}

\end{document}